\documentclass[useAMS,usegraphicx]{mn2e}
\usepackage{bm}

\def\bmr{\bm{r}}
\def\H2{H$_2$}
\def\nH{n_{\rm H}}
\def\fH2{f_{\rm H_2}}
\def\j21{J_{21}}

\begin{document}
\title[H$_2$ in damped Ly$\alpha$ systems]{Molecular hydrogen in
damped Ly$\alpha$ systems: clues to interstellar physics at
high-redshift}
\author[H. Hirashita \& Ferrara]{H. Hirashita$^{1,2}$\thanks{e-mail:
     hirashita@u.phys.nagoya-u.ac.jp}\thanks{Postdoctoral
     Fellow of the Japan Society for the Promotion of
     Science (JSPS).}
and A. Ferrara$^{2}$
\\
$^1$ Graduate School of Science, Nagoya University, Nagoya
     464-8602, Japan \\
$^2$ SISSA/International School for Advanced Studies, Via
     Beirut 4, 34014 Trieste, Italy
}
\date{Accepted 2004 November 8; Received 2004 June 9}
\pubyear{2004} \volume{000} \pagerange{1}
\twocolumn

\maketitle \label{firstpage}
\begin{abstract}
In order to interpret \H2 quasar absorption line observations of damped
Ly$\alpha$ systems (DLAs) and sub-DLAs, we model their \H2 abundance as
a function of dust-to-gas ratio, including \H2 self-shielding and dust
extinction against dissociating photons. Then, we constrain the physical
state of gas by using \H2 data.
Using \H2 excitation data for DLA with \H2 detections, we derive a gas
density $1.5\la\log n~[{\rm cm}^{-3}]\la 2.5$, temperature
$1.5\la\log T~[{\rm K}]\la 3$, 
and internal UV radiation field (in units of the Galactic value)
$0.5\la\log\chi\la 1.5$.  We then find that the observed relation
between molecular fraction 
and dust-to-gas ratio of the sample is naturally explained by the above
conditions. However, it is still possible that  \H2 deficient DLAs and
sub-DLAs with \H2 fractions less than $\sim 10^{-6}$ are in a more
diffuse and warmer state.
The efficient photodissociation by the internal UV radiation field
explains the extremely small \H2 fraction ($\la 10^{-6}$) observed for
$\kappa\la 1/30$ ($\kappa$ is the
dust-to-gas ratio in units of the Galactic value); \H2 self-shielding
causes a rapid increase and the large variations of \H2 abundance for
$\kappa \ga 1/30$. We finally 
propose an independent method to estimate the star formation rates of
DLAs from \H2 abundances; such rates are then critically compared with
those derived from 
other proposed methods. The implications for the  contribution of DLAs
to the cosmic star formation history are briefly discussed.
\end{abstract}

\section{Introduction}

Damped Ly$\alpha$ clouds (DLAs) are quasar (QSO) absorption
line systems whose neutral hydrogen column density is larger
than $\sim 1-2\times 10^{20}$ cm$^{-2}$
(e.g.\ Prochaska \& Wolfe 2002).
DLAs absorb Ly$\alpha$ photons at the restframe of the DLAs.
Because QSOs are generally luminous, DLAs provide us with
unique opportunities to trace high-redshift (high-$z$)
galaxy evolution. In particular, by identifying absorption
lines of various species, the physical condition of
the interstellar medium (ISM) of DLAs has been deduced.

{}From the analyses of various absorption lines, evidence
has been found for the existence of heavy elements in DLAs
(e.g.\ Lu et al.\ 1996). The evolution of metal abundance in
DLAs can trace the chemical enrichment history of present
galaxies. Based on this, and on other clues, DLAs have been
suggested to be the progenitors of nearby galaxies; the
similar values of the baryonic mass density in
DLAs around redshift $z\sim 2$ and the stellar mass density at
$z\sim 0$ has further supported this idea (Lanzetta, Wolfe,
\& Turnshek 1995; Storrie-Lombardi \& Wolfe 2000).
By adopting a recently favoured $\Lambda$CDM cosmology,
however, P\'{e}roux et al.\ (2003) argue that the comoving
density of H {\sc i} gas at $z\sim 2$ is smaller than the
comoving stellar mass density at $z\sim 0$. Yet
they strengthen the importance of DLAs,
showing that a large fraction of
H {\sc i} gas is contained in DLAs at $z\sim 2$--3.

In general, a certain fraction of metals condenses onto
dust grains. Indeed, Fall, Pei, \& McMahon (1989) have
suggested that the reddening of background quasars indicates
typical dust-to-gas ratios of $\sim 1/20$--1/4 of the Milky Way
(see also Zuo et al.\ 1997, but see Murphy \& Liske 2004).
The depletion of heavy elements also supports the dust content
in DLAs (Pettini et al.\ 1994; Vladilo 2002). The existence of
dust implies the possibility that the formation of hydrogen
molecules (\H2) is enhanced because of the \H2 grain surface
reaction (Lanzetta, Wolfe, \& Turnshek 1989).
Hirashita \& Ferrara (2002) argue that the enhancement of
molecular abundance results in an enhancement of the star
formation activity in the early evolutionary stages of galaxy
evolution, because stars form in molecular clouds. The
important role of dust
on the enhancement of the \H2 abundance is also suggested by
observations of DLAs (Ge, Bechtold, \& Kulkarni 2001;
Ledoux, Petitjean, \& Srianand 2003)
and in the local Universe, e.g.\ in Galactic (Milky Way) halo
clouds (e.g.\ Richter et al.\ 2003) and in the Magellanic
Clouds (Richter 2000; Tumlinson et al.\ 2002).

Although the \H2 fraction (the fraction of hydrogen nuclei
in the form of \H2; see equation \ref{eq:H2frac})
is largely enhanced for some DLAs, stringent upper limits are
laid on a significant fraction of DLAs in the range
$\la 10^{-7}$--$10^{-5}$ (Black, Chaffee, \& Foltz 1987;
Petitjean, Srianand, \& Ledoux 2000). This can be interpreted
as due to a low formation rate of \H2 in dust-poor environments
relative to the Milky Way (Levshakov et al.\ 2002; Liszt 2002)
and high \H2 dissociation rate by strong ultraviolet (UV)
radiation
(e.g.\ Petitjean et al.\ 2000). However, we should keep in
mind that such upper limits do not exclude the existence of
molecule-rich clouds in these systems, because molecular
clouds may have a very low volume filling factor. Indeed,
based on the hydrodynamical simulation of
Wada \& Norman (2001),
Hirashita et al.\ (2003) show that under a strong UV field
typical of high-$z$ and a poor dust content
($\sim 1/10$ of the Galactic dust-to-gas ratio), \H2-rich
regions are located in very clumpy small regions. In such a
situation, it is natural that molecular clouds are hardly
detected in DLAs.

The probability of detecting \H2 is higher for DLAs with
larger dust-to-gas ratio or larger metallicity.
Indeed, \H2 tends to be detected for metal-rich DLAs
(Ledoux et al.\ 2003). The correlation between
dust-to-gas ratio and \H2 abundance for DLAs 
indicates that \H2 predominantly forms on dust
grains as in the Galaxy (e.g.\ Jura 1974). Since the
\H2 formation and destruction rates are affected mainly
by gas density, dust-to-gas ratio, and UV radiation
intensity, we can derive or constrain those quantities
for DLAs based on \H2 abundance. Those quantities
also enable us to draw conclusions about other important
quantities such as cooling and heating rates, star
formation rate (SFR), etc.\
(Wolfe, Prochaska, \& Gawiser 2003a).

Our main aim in this paper is to investigate what we can
learn from the recent \H2 observations of DLAs and
sub-DLAs. (Although we mainly focus on DLAs, we also
include low column density systems, which are called
sub-DLAs.) In particular, we focus on key quantities for
\H2 formation and destruction (i.e.\ dust-to-gas ratio
and UV intensity, respectively). The physical state of
\H2 reflects the physical state of gas, especially, gas
density and temperature. In this work, we concentrate on
\H2 data to derive those physical quantities. Recent
{\it FUSE} (Far Ultraviolet Spectroscopic Explorer)
observations of the
Galactic ISM (e.g.\ Gry et al.\ 2002;
Marggraf, Bluhm, \& de Boer 2004),
the Galactic halo clouds (e.g.\ Richter et al.\ 2003),
and the Magellanic Clouds (e.g.\ Tumlinson et al.\ 2002;
Richter, Sembach, \& Howk 2003;
Andr\'{e} et al.\ 2004)
are successful in deriving the physical
state of gas from \H2 absorption line data.
We focus on DLAs and sub-DLAs to investigate the
high-$z$ universe.
Since recent observations
suggest that the local UV radiation originating from star
formation within DLAs is stronger than the
UV background intensity (Ledoux, Srianand, \& Petitjean 2002;
Wolfe et al.\ 2003a), we include the local UV field
in this work. We call the local UV field ``interstellar
radiation field (ISRF)''.

Since it
is still unclear whether DLAs are large protogalactic discs
(Wolfe et al.\ 1986; Prochaska \& Wolfe 1998;
Salucci \& Persic 1999),
small sub-$L^*$ galaxies (Gardner et al.\ 2001;
M{\o}ller et al.\ 2002;
Okoshi et al.\ 2004), protogalactic
clumps (Haehnelt, Steinmetz, \& Rauch 1998;
Ledoux et al.\ 1998), or a mixture
of various populations (Burbidge et al.\ 1996;
Cen et al.\ 2003;
Rao et al.\ 2003), we adopt a simple model which 
nevertheless includes all relevant physical processes 
and derive robust general conclusions. 

We first describe the model we use to derive the molecular
content in DLAs
(Section \ref{sec:model}). After describing
the observational sample adopted in this paper
(Section \ref{sec:sample}), we compare our results with the
data and constrain the physical conditions of DLAs
(Section \ref{sec:results}).
Based on these results,
we extend our discussion to the SFR (Section \ref{sec:sfr}),
and we finally give a summary of this paper
(Section \ref{sec:sum}).

\section{Model}\label{sec:model}

Our aim is to investigate the physical conditions in the ISM of
DLAs by treating \H2 formation and destruction for a
statistical sample. For the homogeneity of our analysis,
we concentrate our interest on \H2. Our aim is not to
analyse the data of each object in detail by using
various absorption lines of various species, since different
lines may originate from
different places and detected lines are different from
object to object. Our model is analytical for the simple
application to a large statistical sample.
There are a lot of works that treat details of various gas
state by fitting the observational results of absorption
lines of various atoms and molecules. Therefore, our model
may be too simple to derive a precise physical quantities
for {\it each} object. However, our analysis has the following
advantages for the statistical purpose: (i) A large sample is
treated homogeneously,
since we concentrate only on \H2 in an analytic way. (ii)
The results
directly conclude the statistical properties of DLAs
and are not affected by details and peculiar situations
of each object.

The relevant physical quantities are those concerning the
\H2 formation and destruction, that is, molecular fraction,
dust-to-gas ratio, UV radiation field, and gas density and
temperature. Observationally, the molecular fraction,
the dust-to-gas ratio and
the H {\sc i} column density are relatively well known, but
the UV radiation field, and the gas density and temperature
are poorly constrained. Thus, we first constrain
the reliable ranges in those quantities by reviewing \H2
detected objects. Then, the likelihood of those parameters
are discussed by using a statistical sample.

\subsection{\H2 formation and destruction}\label{subsec:chemi}

For the metallicity range typical of DLAs, we can assume
equilibrium between \H2 formation and destruction, because
the timescale of \H2 formation and destruction is well below
the dynamical timescale (Hirashita et al.\ 2003). We adopt
the formation rate of \H2 per unit volume and time,
$R_{\rm dust}$, by Hollenbach \& McKee (1979) (see also
Hirashita et al.\ 2002)\footnote{In Hirashita et al.\ (2003),
there is a typographic error. (The results are correctly
calculated.) The expression in this paper is correct and $R_1$
in Hirashita et al.\ (2003) is equal to
$R_{\rm dust}n\nH$.}:
\begin{eqnarray}
R_{\rm dust} & = & 4.1\times 10^{-17}S_{\rm d}(T)\left(
\frac{a}{0.1~\mu{\rm m}}\right)^{-1}\left(\frac{{\cal D}}{10^{-2}}
\right)\nonumber\\
& \times & \left(\frac{T}{100~{\rm K}}
\right)^{1/2}\left(\frac{\delta}{2~{\rm g}~{\rm cm}^{-3}}\right)
~{\rm cm^{3}~s^{-1}}\, ,
\label{eq:formation}
\end{eqnarray}
where $a$ is the radius of a grain (assumed to be spherical with
a radius of 0.1 $\mu$m unless otherwise stated), ${\cal D}$ is
the dust-to-gas mass ratio (varied in this paper; the typical
value in the solar vicinity of the Milky Way is $10^{-2}$),
$\delta$ is the grain material density (assumed to be
2 g cm$^{-3}$ in this paper), and
$S_{\rm d}(T)$ is the sticking coefficient of hydrogen atoms onto
dust. The sticking coefficient is given by (Hollenbach \& McKee
1979; Omukai 2000)
\begin{eqnarray}
S_{\rm d}(T) & = & [1+0.04(T+T_{\rm d})^{0.5}+2\times 10^{-3}T+8
\times 10^{-6}T^2]^{-1}\nonumber \\
& \times & [1+\exp(7.5\times 10^2(1/75-1/T_{\rm d}))]^{-1}\, ,
\label{eq:stick}
\end{eqnarray}
where $T_{\rm d}$ is the dust temperature, which is calculated
by assuming the radiative equilibrium in equation (\ref{eq:T_d}).
However, since the reaction rate is insensitive to the dust
temperature as long as $T_{\rm d}\la 70$ K, the following
results are not affected
by the dust temperature. In fact, $T_{\rm d}$ never exceeds
70 K under the radiation field intensity derived in this paper.
The \H2 formation rate per unit volume is estimated
by $R_{\rm dust}n\nH$, where $n$ is the gas number density,
and $\nH$ is the number density of H {\sc i}.

The \H2 formation rate on dust grains is still to be debated.
It depends on grain size (eq.\ \ref{eq:formation}).
If the grain size is much smaller than 0.1 $\mu$m, the \H2
formation rate is largely enhanced. There are uncertainties
also in $S_{\rm d}(T)$, which could depend on the materials of
dust. Cazaux \& Tielens (2002) also suggest that $S_{\rm d}(T)$
in equation (\ref{eq:formation}) should be substituted by
$\epsilon_{\rm H_2}S_{\rm d}(T)$, where $\epsilon_{\rm H_2}$
is the recombination efficiency (the fraction of the
accreted hydrogen that desorbs as \H2). The recombination
efficiency is $\epsilon_{\rm H_2}\sim 1$ if $5\la T\la 30$ K,
and $\epsilon_{\rm H_2}$ becomes $\sim 0.2$ at $T\sim 100$ K.
If the temperature is higher than $\sim 300$ K,
$\epsilon_{\rm H_2}\sim 0$. However, if
$\epsilon_{\rm H_2}\sim 0.2$ is multiplied to the \H2
formation rate at $T=100$ K,
the grain formation rate becomes significantly small, and
the Galactic \H2 formation rate derived for $T\sim 100$ K
by Jura (1974)
cannot be achieved; we have to keep in mind that 
the theoretical determination of $R_{\rm dust}$ is affected by some uncertainty. Thus,
in this paper, we adopt equation (\ref{eq:formation})
to assure that it provides the Galactic \H2 formation rate
if we adopt $a\sim 0.1~\mu$m,  ${\cal D}\sim 0.01$,
$T\sim 100$ K, and $\delta\sim 2$ g cm$^{-3}$.
Since it is important to recongnise that the temperature
dependence of $R_{\rm dust}$ is uncertain, we do not
deeply enter discussions on the temperature dependence
in this paper.

The photodissociation rate $R_{\rm diss}$ is estimated as
(Abel et al.\ 1997)
\begin{eqnarray}
R_{\rm diss}=(4\pi )\, 1.1\times 10^{8}J_{\rm LW}\,
S_{\rm shield}~{\rm s}^{-1}\, ,\label{eq:R_diss}
\label{eq:dissociation}
\end{eqnarray}
where $J_{\rm LW}$ (erg s$^{-1}$ cm$^{-2}$ Hz$^{-1}$ sr$^{-1}$)
is the UV intensity at $h\nu =12.87$ eV averaged over the
solid angle, $S_{\rm shield}$ is the correction factor of the
reaction rate for \H2 self-shielding
and dust extinction. The photodissociation rate per unit
volume is estimated as $R_{\rm diss}n_{\rm H_2}$, where
$n_{\rm H_2}$ is the number density of \H2.
We adopt the correction for the \H2
self-shielding by
Draine \& Bertoldi (1996)\footnote{The exact treatment of
self-shielding slightly deviates from the fitting formula
that we adopt (see Figures 3--5 of
Draine \& Bertoldi 1996; the difference is within a factor
of $\sim 2$). Accordingly, the estimate of radiation field
in Section \ref{subsec:each} is affected by the same amount.
The difference in the shielding factor may cause a large
difference in theoretically calculated molecular fraction
(Abel et al.\ 2004). This point is commented in
the last paragraph of Section \ref{subsec:scaling}.}
(see also
Hirashita \& Ferrara 2002). Then, we estimate
$S_{\rm shield}$ as
\begin{eqnarray}
S_{\rm shield}={\rm min}\left[1,\, \left(
\frac{N_{\rm H_2}}{10^{14}~{\rm cm}^{-2}}\right)^{-0.75}\right]
e^{-\sigma_{\rm d}N_{\rm d}}\, ,
\end{eqnarray}
where $N_{\rm H_2}$ and $N_{\rm d}$ are the column densities of
\H2 and dust, respectively, and $\sigma_{\rm d}$ is the cross
section of a grain against \H2 dissociating photons. In the UV
band, $\sigma_{\rm d}$ is approximately estimated with the
geometrical cross section: $\sigma_{\rm d}\simeq \pi a^2$
(Draine \& Lee 1984). The column density of grains is related
to $N_{\rm H}$:
$(4/3)\pi a^3\delta N_{\rm d}=1.4m_{\rm H}N_{\rm H}{\cal D}$
(the factor 1.4 is the correction for the helium content).
Therefore, the optical depth of dust in UV,
$\tau_{\rm UV}\equiv\sigma_{\rm d}N_{\rm d}$, is
expressed as
\begin{eqnarray}
\tau_{\rm UV} & = &
\frac{4.2N_{\rm H}m_{\rm H}{\cal D}}{4a\delta}\nonumber\\
& = &  0.879\left(\frac{a}{0.1~\mu{\rm m}}\right)^{-1}
\left(\frac{\delta}{2~{\rm g~cm}^{-3}}\right)^{-1}
\left(\frac{{\cal D}}{10^{-2}}\right)\nonumber\\
& & {}\times\left(
\frac{N_{\rm H}}{10^{21}~{\rm cm}^{-2}}\right)\, .
\label{eq:tauUV}
\end{eqnarray}
Since ${\cal D}<10^{-2}$ for DLAs (often
${\cal D}\ll 10^{-2}$), the dust extinction
is negligible except for the high column density
and dust-rich regime satisfying
$N_{\rm H}{\cal D}\ga 10^{19}~{\rm cm}^{-2}$.
We also define the critical molecular fraction,
$\fH2^{\rm cr}$ as
the molecular fraction which yields
$N_{\rm H_2}=10^{14}$ cm$^{-2}$:
\begin{eqnarray}
\fH2^{\rm cr}(N_{\rm H})\equiv 2\times 10^{14}/N_{\rm H}\,
.\label{eq:critical}
\end{eqnarray}
For $\fH2 >\fH2^{\rm cr}$, the self-shielding affects the
\H2 formation.

In this paper, we assume the Galactic (Milky Way)
dust-to-gas ratio to be ${\cal D}_\odot =0.01$. We
define the normalised dust-to-gas ratio $\kappa$ as
\begin{eqnarray}
\kappa\equiv{\cal D}/{\cal D}_\odot\, .
\end{eqnarray}

The typical Galactic ISRF intensity has been estimated to
be $c\nu u_\nu =1.2\times 10^{-3}$ erg cm$^{-2}$ s$^{-1}$
at the wavelength of 1000 \AA\ (i.e.\
$\nu =3.0\times 10^{15}$ Hz), where $u_\nu$ is the radiation
energy density per unit frequency (Habing 1968).
Approximating the energy density of
photons at 1000 \AA\ with that at the Lyman-Werner Band,
we obtain for $J_{\rm LW}$ at the solar vicinity,
$J_{\rm LW\odot}\simeq cu_\nu /4\pi =3.2\times 10^{-20}$
erg s $^{-1}$ cm$^{-2}$ Hz$^{-1}$ sr$^{-1}$.
The intensity normalised to the Galactic ISRF, $\chi$, is
defined by
\begin{eqnarray}
\chi\equiv J_{\rm LW}/ J_{\rm LW\odot}\, .\label{eq:chi}
\end{eqnarray}
Using equations (\ref{eq:R_diss}) and (\ref{eq:chi}),
we obtain
\begin{eqnarray}
R_{\rm diss}=4.4\times 10^{-11}\chi
S_{\rm shield}~{\rm s}^{-1}\, .\label{eq:diss}
\end{eqnarray}
With $\chi =1$, the formula reproduces the typical Galactic
photodissociation rate (2--$5\times 10^{-11}~{\rm s}^{-1}$)
derived by Jura (1974).

The UV background intensity at the Lyman limit is estimated
to be $J_{21}=0.3$--1 around $z\sim 3$, where $J_{\rm 21}$ is
in units of $10^{-21}$ erg cm$^{-2}$ s$^{-1}$ Hz$^{-1}$
sr$^{-1}$ (Haardt \& Madau 1996; Giallongo et al.\ 1996;
Cooke, Espey, \& Carswell 1997; Scott et al.\ 2000;
Bianchi, Cristiani, \& Kim 2001). If we assume that
the Lyman limit intensity is equal to the Lyman-Werner
luminosity, we roughly obtain $\chi =32J_{\rm 21}$.
The UV background intensity is
likely to be lower at $z\la 1$ (e.g.\ Scott et al.\ 2002).
Thus, the internal stellar radiation is dominant
for \H2 dissociation if $\chi\ga 1/32$.

As mentioned at the beginning of this subsection, we can
assume that the formation and destruction of \H2 are
in equilibrium. Therefore, the following equation holds:
\begin{eqnarray}
R_{\rm dust}n\nH =R_{\rm diss}n_{\rm H_2}\, .
\label{eq:equil}
\end{eqnarray}
In order to solve this equation, it is necessary to give
the gas temperature $T$, the gas number density $n$, the
normalised dust-to-gas ratio $\kappa$, the normalised
ISRF $\chi$, and the column density $N_{\rm H}$.
As for the dust grains,
we assume $a=0.1~\mu$m, and
$\delta =2~{\rm g}~{\rm cm}^{-3}$, all of which are
typical values in the local universe.
After solving that equation, the molecular fraction is
obtained from the definition:
\begin{eqnarray}
f_{\rm H_2}\equiv\frac{2n_{\rm H_2}}{\nH +2n_{\rm H_2}}
\, ,\label{eq:H2frac}
\end{eqnarray}
where we neglect the ionised hydrogen for DLAs.

\subsection{Dust temperature}

Dust temperature is necessary to calculate the sticking
efficiency of H onto grains (equation \ref{eq:stick}).
The equilibrium dust temperature is determined from the
balance between incident radiative energy and radiative
cooling. We adopt the equilibrium temperature calculated
by Hirashita \& Hunt (2004) (see also
Takeuchi et al.\ 2003):
\begin{eqnarray}
T_{\rm d} & = & 12\, (\chi Q_{\rm UV})^{1/6}\left(
\frac{A}{3.2\times 10^{-3}~{\rm cm}}\right)^{-1/6}\nonumber\\
 & & {}\times\left(
\frac{a}{0.1~\mu{\rm m}}\right)^{-1/6}~{\rm K}\, ,
\label{eq:T_d}
\end{eqnarray}
where the constant $A$ depends on the optical properties of dust grains, and
for silicate grains $A=1.34\times 10^{-3}$ cm (Drapatz \& Michel
1977), while for carbonaceous grains $A=3.20\times 10^{-3}$ cm
(Draine \& Lee 1984; Takeuchi et al.\ 2003). We hereafter
assume $Q_{\rm UV}=1$, $A=3.20\times 10^{-3}$ cm, and
$a=0.1~\mu$m; these assumptions affect very weakly our calculations 
due to the 1/6 power index dependence of those parameters.

\subsection{Approximate scaling of molecular fraction}
\label{subsec:scaling}

As a summary of our formulation, we derive an approximate
scaling relation for $f_{\rm H_2}$. The molecular fraction
is determined from equation (\ref{eq:equil}). The
left-hand side describes the formation rate, which follows
the scaling relation,
$R_{\rm dust}\propto \kappa T^{1/2}S_{\rm d}(T)$.
On the other hand, the \H2 destruction coefficient
$R_{\rm diss}$ is proportional to $\chi$ for
$\fH2 N_{\rm H}<10^{14}~{\rm cm}^{-2}$ ({\it unshielded
regime}), and is proportional to
$\chi (\fH2 N_{\rm H})^{-0.75}$
for $\fH2 N_{\rm H}>10^{14}~{\rm cm}^{-2}$ ({\it shielded
regime}). As mentioned in Section \ref{subsec:chemi}, the
self-shielding is more important than the dust
extinction. Thus, we neglect the
dust extinction in this subsection. 

Under the condition that $n_{\rm H_2}\ll n_{\rm H}$,
$\fH2\simeq 2n_{\rm H_2}/n_{\rm H}\ll 1$,
and by using the expressions above for $R_{\rm dust}$ and
$R_{\rm diss}$, we obtain the following scaling relations
for $\fH2$ from equation (\ref{eq:equil}):
in the unshielded regime
($\fH2 N_{\rm H}<10^{14}~{\rm cm}^{-2}$),
\begin{eqnarray}
f_{\rm H_2}\propto\kappa n\chi^{-1}T^{1/2}
S_{\rm d}(T)\, ,
\end{eqnarray}
and in the shielded regime
($\fH2 N_{\rm H}>10^{14}~{\rm cm}^{-2}$),
\begin{eqnarray}
f_{\rm H_2}\propto [\kappa n\chi^{-1}T^{1/2}
S_{\rm d}(T)]^4N_{\rm H}^3\, .
\end{eqnarray}

We have adopted the approximate formula
$S_{\rm shield}\propto N_{\rm H_2}^{-0.75}$
for $N_{\rm H_2}>10^{14}~{\rm cm}^{-2}$.
In fact, there is a slight (factor of $\sim 2$)
difference between this scaling and the exact
calculation (Figs.\ 3--5 of
Draine \& Bertoldi 1996). In the self-shielding
regime, this slight difference may cause an order
of magnitude difference in the calculated $\fH2$
because of
a nonlinear dependence on $N_{\rm H_2}$
(see the appendix of Abel et al.\ 2004).
In practice, this difference can also be examined
by changing $R_{\rm dust}$ in stead of
$R_{\rm diss}$, and indeed, if we change
$R_{\rm dust}$ by a factor of $\sim 3$, the
calculated molecular fraction is significantly
affected as shown in Section \ref{sec:results}.
For the statistical purpose, however, our
conclusions are robust, since
$N_{\rm H_2}$ spans over a wide range in our sample
and the fitting formula approximate the overall
trend of $S_{\rm shield}$
as a function of $N_{\rm H_2}$ very well.

\section{DATA SAMPLE}\label{sec:sample}

\subsection{Overview}

Recently, Ledoux et al.\ (2003) have compiled a sample of
Ly$\alpha$ clouds with \H2 observations. The H {\sc i}
column density of their sample ranges from
$\log N(\mbox{H {\sc i}})=19.35$ to 21.70. They have
found no correlation between $\log N(\mbox{H {\sc i}})$
and $\log f_{\rm H_2}$, but have found clear correlation
between the metal depletion (or dust-to-gas ratio,
$\log\kappa$) and $\log f_{\rm H_2}$. Although some
clouds have H {\sc i} column densities smaller than the
threshold for DLAs (typically
$\log N(\mbox{H {\sc i}})\ga 20.3$), we treat all the
sample, because there is no physical reason for
using this threshold for H {\sc i}. Such low-column density
objects called sub-DLAs can be useful to
investigate a wider range of $N(\mbox{H {\sc i}})$ and get an  
an insight into several
physical processes, especially self-shielding. For the
absorption system of Q 0013$-$004
(the absorption redshift is $z_{\rm abs}=1.973$),
we adopt the mean of the two values, $-1.73$, for
$\log f_{\rm H_2}$ (but in the figures, the observationally
permitted range, $-2.81<\log f_{\rm H_2}<-0.64$, is shown
by a bar).
Since it is likely that almost all the hydrogen
nuclei are in the form of H {\sc i}, we use
$N(\mbox{H {\sc i}})$ and $N_{\rm H}$ interchangeably.
(If the ionised hydrogen is not negligible in a system,
we can interpret our result to be representative of the
neutral component in the system.)

Ledoux et al.\ (2003) observationally estimate the dust-to-gas
ratio $\kappa$ from the metal depletion:
\begin{eqnarray}
\kappa =10^{[{\rm X/H}]}(1-10^{[{\rm Fe/X}]})\, ,
\label{eq:kappa_obs}
\end{eqnarray}
where X represents a reference element that is little affected
by dust depletion effects. We adopt this formula in this paper.
A formal derivation of equation (\ref{eq:kappa_obs}) is given
by Wolfe et al.\ (2003a). Ledoux et al.\ (2003) have shown a
correlation between
$\log\kappa$ and $\log\fH2$, which strongly suggests that \H2
forms on dust grains. Stringent upper limits are laid on DLAs
with $\log\kappa\la -1.5$. It should also be noted that
for DLAs with $\log\kappa\ga -1.5$ there is a large
dispersion in $\log f_{\rm H_2}$. This dispersion implies that
the molecular abundance
is not determined solely by the dust-to-gas ratio.
Therefore, it is necessary to model
the statistical dispersion of $\log f_{\rm H_2}$.

\begin{figure}
\includegraphics[width=8cm]{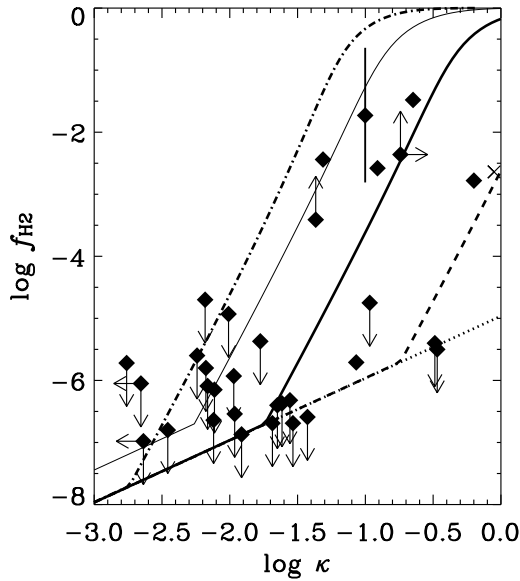}
\caption{
Relation between $f_{\rm H_2}$ (the molecular fraction
of hydrogen) and $\kappa$ (dust-to-gas ratio normalised by
the Galactic value). The squares are the observational data
compiled by Ledoux et al.\ (2003). The data with upward and
downward arrows indicate that only lower and upper limits
for $f_{\rm H_2}$ are obtained, respectively, and those
with rightward and leftward arrows
show lower and upper limits for $\kappa$, respectively.
The cross is the data of Reimers et al.\ (2003).
The thick solid, dotted, dashed, dot-dashed lines present
our calculations for $n=100~{\rm cm}^{-3}$,
$T=100$ K, and $\chi =10$ with different H {\sc i}
column densities ($10^{21}$, $10^{19}$, $10^{20}$, and
$10^{22}$ cm$^{-2}$, respectively). The thin solid line
represents the calculation with the same parameters
as the thick solid line except for a high formation rate
of $R_{\rm dust}$
(small grains with $a=0.03~\mu$m is assumed and
$R_{\rm dust}$ becomes 3.3 times larger).
\label{fig:h2_dust}}
\end{figure}

In Figure \ref{fig:h2_dust}, we plot the relation between
molecular fraction and dust-to-gas ratio of
Ledoux et al.\ (2003) (see their paper for details and
error bars). We also plot the data of a Ly$\alpha$ absorber
in which \H2 is detected (the absorption redshift,
$z_{\rm abs}=1.15$) toward the QSO HE 0515$-$4414
(Reimers et al.\ 2003). For this absorber, we adopt
$\log N(\mbox{H {\sc i}})=19.88$,
$\log f_{\rm H_2}=-2.64$, and $\log\kappa =-0.051$
(Reimers et al.\ 2003). The lines in those figures are our
model predictions, which
are explained in the following section.

We also use the H {\sc i} column density. In
particular, Ledoux et al.\ (2003) show that there is no
evidence of correlation between H {\sc i} column density
and molecular fraction.

\subsection{Individual \H2-detected objects}\label{subsec:each}

In order to derive the reasonable range of physical
quantities characterising Ly$\alpha$ absorbers, we use the 
\H2 data. The objects whose \H2 absorption lines are
detected allow us to constrain the
physical state of gas. As mentioned in
Section \ref{sec:model}, our analysis is limited to \H2.
The excitation temperature is related to the
ratio of the column densities as
(e.g.\ Levshakov et al.\ 2000):
\begin{eqnarray}
\frac{N(J)}{N(0)}=\frac{g_J}{g_0}\exp\left[ -
\frac{B_vJ(J+1)}{T_{J0}}\right]\, ,
\end{eqnarray}
where $N(J')$ is the column density of \H2 in the
level $J=J'$, the statistical weight of a level $J$,
$g_J$, is $3(2J+1)$ for odd $J$ and $(2J+1)$ for even $J$,
and the
constant $B_v=85.36$ K is applicable to the
vibrational ground state. If the column densities of
\H2 in the rotational $J$-th and ground states are
known, we can determine
the excitation temperature $T_{J0}$ by solving the
above equation. In particular, $T_{10}$ is the best
tracer for the kinetic temperature (Jura 1975).
Therefore, we approximate
$T\sim T_{10}$ in the discussion of this section.

Using the column densities of $J=4$ or 5 level, we
can constrain the gas density following
the simple procedure introduced by Jura (1975).
Those levels are populated by direct
formation into these levels
of newly created molecules, and by pumping from
$J=0$ and $J=1$. The assumption that the
lowest two levels ($J=0$ and 1) are dominated in
population holds also to the \H2 detected DLAs.
Following Jura (1975) we adopt  $\beta(0)\simeq\beta (1)$
(see also Jura 1974), where $\beta (J)$ is the
total rate of an upward radiative transition from
level $J$ by absorbing Lyman- and Werner-band photons.
This assumption is true as long as the saturation
levels of absorption are the same. The distribution
function of the levels of formed \H2 is taken from
Spitzer \& Zweibel (1974), who treat a typical
interstellar condition.
We use the ratio of two column densities at the levels
$J=1$ and $J=0$, $N(1)/N(0)$, and assume that
$f_{1,0}\equiv n_{\rm H_2}(1)/n_{\rm H_2}(0)=N(1)/N(0)$,
where $n_{\rm H_2}(J')$ is the hydrogen number density in
the level $J=J'$.
Then we can rewrite equations (3a) and (3b) of
Jura (1975) in the following forms:
\begin{eqnarray}
A(4\to 2)\frac{n_{\rm H_2}(4)}{\nH}=R_{\rm dust}n\left(
0.19+\frac{9.1}{1+f_{1,0}}p_{4,0}\right)\, ,
\label{eq:jura3a}\\
A(5\to 3)\frac{n_{\rm H_2}(5)}{\nH}=R_{\rm dust}n\left(
0.44+\frac{9.1f_{1,0}}{1+f_{1,0}}p_{5,1}\right)\, ,
\label{eq:jura3b}
\end{eqnarray}
where $p_{4,0}$ and $p_{5,1}$ are the pumping
efficiencies into the $J=4$ and $J=5$ levels from
the $J=0$ and $J=1$ levels, respectively, and
$A(J'\to J'')$
denotes the spontaneous transition
probability from $J=J'$ to $J=J''$.
The \H2 formation rate coefficient
$R_{\rm dust}$, the gas particle number density
$n$, and the number density of H {\sc i}
$\nH$ are defined in Section \ref{subsec:chemi}.
We adopt
$A(4\to 2)=2.76\times 10^{-9}~{\rm s}^{-1}$,
$A(5\to 3)=9.85\times 10^{-9}~{\rm s}^{-1}$
(Spitzer 1978), $p_{4,0}=0.26$, and
$p_{5,1}=0.12$ (Jura 1975).

Assuming $n_{\rm H_2}(4)/\nH =N(4)/N_{\rm H}$ and
$n_{\rm H_2}(5)/\nH =N(5)/N_{\rm H}$, we estimate
the gas number density $n$ from
equations (\ref{eq:jura3a}) and/or
(\ref{eq:jura3b}). $R_{\rm dust}$ is given by
equation (\ref{eq:formation}), where we adopt the
observational dust-to-gas ratio for $\kappa$ and
the excitation temperature $T_{10}$ for $T$.
Here we assume that $T_{\rm dust}=12$ K, because
we do not know the ISRF at this stage. However,
this assumption does not affect the result in
the range of $T_{\rm dust}$ consistent with the
range of $\chi$ estimated below.

Finally equation (\ref{eq:equil}) is used to obtain
the UV radiation field $\chi$ (proportional to
$R_{\rm diss}$; equation \ref{eq:diss}, where
$S_{\rm shield}$ depends on $N_{\rm H_2}$) by using
the estimated $T=T_{10}$ and $n$, and the observed
value of
$n_{\rm H_2}/\nH =N_{\rm H_2}/N_{\rm H}$.
Each \H2-detected object is discussed separately in the
following. This simple analytical method suffices for the
statistical character of our study.  Careful treatments
focusing on individual objects may require
a more detailed treatment of \H2 excitation state
as described, for example, in
Browning, Tumlinson, \& Shull (2003).

\subsubsection{Q 0013$-$004 ($z_{\rm abs}=1.968$)}

The excitation temperature is estimated to be
$T_{10}=73$ K (Petitjean, Srianand, \& Ledoux 2002). Only the
upper limit is obtained for $N_{\rm H}$
($\log N_{\rm H}~[{\rm cm}^{-2}]\leq 19.43$).
Thus, $N(4)/N_{\rm H}\geq 3.2\times 10^{-5}$
and $N(5)/N_{\rm H}\geq 1.3\times 10^{-5}$.
Those two values indicate
$R_{\rm dust}n\geq 6.0\times 10^{-14}~{\rm s}^{-1}$
and
$R_{\rm dust}n\geq 1.4\times 10^{-13}~{\rm s}^{-1}$,
respectively. This formation rate is very high
compared with the Galactic molecular clouds
(e.g.\ Jura 1975), and if the Galactic
dust-to-gas ratio is assumed, we obtain
$n>6\times 10^3$ cm$^{-3}$. However, with this
high density, the levels $J=2$ should be in thermal
equilibrium, but $T_{20}=302$ K, which is higher
than $T_{10}$. This indicates that the gas density
should be less than the critical density,
$n\la 200~{\rm cm}^{-3}$, which is consistent with
the density derived from C {\sc i} lines
(10--85 cm$^{-3}$; Petitjean et al.\ 2002). The
discrepancy between a density derived from \H2 and
one from C {\sc i} is also reported for the
absorption system of HE 0515$-$4414
(Reimers et al.\ 2003; Section \ref{subsubsec:he}).
They suspect that the \H2 formation rate is larger
than the Galactic value. For example, if the grain
size is typically smaller than that assumed in
equation (\ref{eq:formation}),
the formation rate becomes larger.

Since only the lower limits are obtained for
$N_{\rm H_2}/N_{\rm H}$ and $R_{\rm dust}n$, it is
not possible to determine $\chi$. By using
$\log N_{\rm H_2}=16.77$, we obtain
$S_{\rm shield}=8.4\times 10^{-3}$.
It may be reasonable to assume that
$N_{\rm H}/N_{\rm H_2}\ga 1$, and in this case,
we obtain $R_{\rm diss}=R_{\rm dust}n
(N_{\rm H}/N_{\rm H_2})\ga 6.0\times 10^{-14}$
s$^{-1}$. With this $R_{\rm diss}$ and the above
$S_{\rm shield}$, we obtain
$\chi\ga 0.16$, supporting the
existence of radiation field whose intensity is
roughly comparable to or larger than the
Galactic one.

\subsubsection{Q 0013$-$004 ($z_{\rm abs}=1.973$)}

The \H2 abundance is only poorly constrained, and it is
impossible to determine the excitation temperature.
We assume the same excitation temperature as the
previous object $z_{\rm abs}=1.968$ ($T_{10}=73$ K).
The excitation state $N(4)/N_{\rm H}=3.2\times 10^{-5}$
is interpreted to be
$R_{\rm dust}n=4.1\times 10^{-15}~{\rm s}^{-1}$. This
object has $\kappa =0.099$, and thus,
$R_{\rm dust}=2.2\times 10^{-18}~{\rm cm}^3~{\rm s}^{-1}$.
Therefore, we again obtain a high density
$n\sim 2000~{\rm cm}^{-3}$.
For this object, there is a large observationally
permitted range of $N_{\rm H_2}$
($S_{\rm shield}=3.0\times 10^{-5}$--$1.5\times 10^{-3}$),
and correspondingly
the range of $\chi =23$--80 is derived.

\subsubsection{Q 0347$-$383 ($z_{\rm abs}=3.025$)}

The $J=1$ level is highly populated relative to
$J=0$, which suggests that the kinetic temperature
of this system is very high ($T_{10}\ga 880$ K;
Ledoux et al.\ 2003). Levshakov et al.\ (2002)
show that the excitation temperature of 825 K is
applicable to the $J=0$--5 levels. However, they
also show that
the width of the \H2 lines indicate that the kinetic
temperature is less than 430 K. The excitation
state $N(4)/N_{\rm H}=3.6\times 10^{-8}$
indicate that $R_{\rm dust}n=2\times 10^{-16}
~{\rm s}^{-1}$, consistent with
Levshakov et al.\ (2002). With $\kappa =0.0857$,
we obtain
$R_{\rm dust}=1.8\times 10^{-18}~{\rm cm}^{-3}
~{\rm s}^{-1}$ and
$1.1\times 10^{-18}~{\rm cm}^{-3}~{\rm s}^{-1}$
for $T=400$ K and 800 K, respectively.
Thus, we obtain $n=100$--200 cm$^{-3}$.
Levshakov et al.\ (2002) derive a gas density much
lower (6 cm$^{-3}$) than these, but they assume
much higher
$R_{\rm dust}$ comparable to the Galactic value.
However, such a large $R_{\rm dust}$ is difficult
to achieve, since the dust-to-gas
ratio is much smaller. The UV radiation field
derived from $R_{\rm dust}n$ and
$S_{\rm shield}=0.38$ is $\chi =15$.

\subsubsection{Q 0405$-$443 ($z_{\rm abs}=2.595$)}

The excitation temperature is $T_{10}=100$ K
(Ledoux et al.\ 2003). The upper limit for $N(4)$ is
obtained ($N(4)/N_{\rm H}<9.3\times 10^{-8}$).
This upper limit is interpreted to be
$R_{\rm dust}n<2.4\times 10^{-16}~{\rm s}^{-1}$,
and using
$R_{\rm dust}=1.2\times 10^{-18}~{\rm cm}^3~
{\rm s}^{-1}$ derived from $\kappa =0.049$, we
obtain $n<200~{\rm cm}^{-3}$.
The upper limit of the radiation field is estimated
form the upper limit of $R_{\rm dust}n$ as
$\chi <4.1$ (with $S_{\rm shield}=7.3\times 10^{-4}$).

\subsubsection{Q 0528$-$250 ($z_{\rm abs}=2.811$)}

There is no information on the excitation state of
\H2.

\subsubsection{Q 0551$-$366 ($z_{\rm abs}=1.962$)}

The component $z_{\rm abs}=1.96214$ dominates the
\H2 content in this system (Ledoux et al.\ 2002).
For this component, the excitation temperature is
$T_{10}=110$ K. From
$N(4)/N_{\rm H}=7.1\times 10^{-7}$, we obtain
$R_{\rm dust}n=2.0\times 10^{-15}~{\rm s}^{-1}$,
and with $R_{\rm dust}=1.6\times 10^{-17}$
($\kappa =0.63$), we obtain $n=130~{\rm cm}^{-3}$.
The shielding factor is
$S_{\rm shield}=2.3\times 10^{-3}$, which indicates
that the UV radiation field is estimated to be
$\chi =24$. The C {\sc i} fine structure levels
indicate $n\sim 55$--390 cm$^{-3}$, consistent with
our estimate.

\subsubsection{Q 1232+082 ($z_{\rm abs}=2.338$)}

This object is observed by
Srianand, Petitjean, \& Ledoux (2000). The
excitation temperature is $T_{10}=185$ K.
Varshalovich et al.\ (2001) observe HD lines
and estimate the excitation temperature from
the ratio of $J=1$ and $J=0$ populations as
$T=70\pm 7$ K. This is lower than above, but
the cold gas phase is supported.

The fraction of the $J=4$ population is
$N(4)/N_{\rm H}=5.6\times 10^{-7}$, which leads
to
$R_{\rm dust}n=2.2\times 10^{-15}~{\rm s}^{-1}$.
By using $\kappa =0.043$, we obtain
$R_{\rm dust}=1.1\times 10^{-18}~{\rm cm}^3
~{\rm s}^{-1}$. Then, the density is estimated
to be $n=2000~{\rm cm}^{-3}$. We can also use
$N(5)/N_{\rm H}=5.8\times 10^{-7}$ to derive
$n=4100~{\rm cm}^{-3}$. Those densities are large
enough to thermalise the level $J=2$, but this
level is not thermalised. Therefore,
Srianand et al.\ (2000) argue that $n$ should
be much smaller. They derive the gas density
$20<n_{\rm H}<35~{\rm cm}^{-3}$ from the
C {\sc i} observation. The absorption systems
of Q 0013$-$004 and HE 0515$-$4414 also
have the same problem that the density
derived from \H2 excitation is too high.
The \H2 formation rate $R_{\rm dust}$ may be
larger than that estimated here probably because
of a small grain size.
Ledoux et al.\ (2003) argue that the molecular
fraction $\log\fH2 =-3.41$
($\log N_{\rm H_2}=17.19$) should be taken as
a lower limit. Therefore,
$S_{\rm shield}\geq 3.9\times 10^{-3}$, and
an upper limit for the
UV radiation field is estimated to be
$\chi\leq 64$.

\subsubsection{Q 1444+014 ($z_{\rm abs}=2.087$)}

The excitation measurements indicate that
$T_{10}=190$ K (Ledoux et al.\ 2003). From the
observational upper limit of the $J=4$ population,
$N(4)/N_{\rm H}\leq 1.4\times 10^{-6}$, we obtain
$R_{\rm dust}n\leq 5.4\times 10^{-15}~{\rm s}^{-1}$
by using $\kappa =0.225$. Then we obtain
$n\leq 950~{\rm cm}^{-3}$. With
$S_{\rm shield}=5.8\times 10^{-4}$, the
upper limit of $R_{\rm dust}$ gives the upper limit
of the ISRF of
$R_{\rm dust}n$ is $\chi\leq 12$.

\subsubsection{HE 0515$-$4414 ($z_{\rm abs}=1.15$)}
\label{subsubsec:he}

This object has been observed by
Reimers et al.\ (2003). The excitation temperature
$T_{10}=90$ K and the excitation state
$N(4)/N_{\rm H}=1.3\times 10^{-5}$ are obtained.
Then, we obtain
$R_{\rm dust}n=3.1\times 10^{-14}~{\rm s}^{-1}$.
By using $\kappa =0.89$, we estimate that
$R_{\rm dust}=2.1\times 10^{-17}~{\rm cm}^3
~{\rm s}^{-1}$, and the density is thus estimated
to be 1500 cm$^{-3}$. Again we obtain a very high
density, but the $J=2$ level is not thermalised
(Reimers et al.\ 2003). Therefore, the density
should be much lower. Reimers et al.\ (2003) also
report the same
problem, adopting the density derived
from C {\sc i} ($n\sim 100~{\rm cm}^{-3}$;
Quast, Baade, \& Reimers 2002).
Perhaps a large \H2 formation rate
$R_{\rm dust}$ is required as in Q 0013$-$004
($z_{\rm abs}=1.968$) and Q 1232+082
($z_{\rm abs}=2.338$). With
$S_{\rm shield}=5.9\times 10^{-3}$, the ISRF
estimated from $R_{\rm dust}n$ is $\chi =100$.

\subsection{Typical physical state}
\label{subsec:typical}

Although the
\H2 detected objects have a wide range of
physical quantities, they are roughly consistent
with a cold phase whose typical density and
temperature are $n\sim 100~{\rm cm}^{-3}$,
$T\sim 100$ K, respectively. The ISRF is
generally larger than the Galactic value, and
the intensity is roughly summarised as an
order-of-magnitude difference, $\chi\sim 10$.
Some objects may indicate much higher densities
($n\sim 1000~{\rm cm}^{-3}$) than those derived
from C {\sc i} excitation states. The
discrepancy may be due to the high \H2
formation rate on dust grains, suggesting that
$R_{\rm dust}$ may be larger than that
assumed in this paper. In view of
equation (\ref{eq:formation}), $R_{\rm dust}$
becomes larger if the typical size of grains
is smaller.

Our previous work (Hirashita et al.\ 2003)
suggests that the covering fraction
of the region with a density larger than
1000 cm$^{-3}$ is low ($<1$\%) because of
small sizes of such dense regions.
Indeed if we assume the density of
1000 cm$^{-3}$ and the column density of
$10^{21}~{\rm cm}^{-2}$, we obtain the size of
0.3 pc. Because of small geometrical cross
sections, such dense regions rarely
exist in the line of sight of QSOs. Therefore,
for the statistical purpose, we do not
go into details of very dense
($\ga 1000~{\rm cm}^{-3}$) regions.

The extremely high density might be an artefact
of our one-zone treatment. If the destruction of
\H2 occurs selectively on the surface of the clouds
and the formation of \H2 occurs in the interior of
the clouds, our approach will not work. However,
the formation and destruction
should be balanced globally, and our approach
could give the first approximation of such a
global equilibrium.

We should note that those quantities are derived
only from the \H2 detected objects. Those
could be associated with the star-forming
regions, since stars form in molecular clouds.
Therefore, the physical quantities derived
from \H2-detected objects could be biased
to high radiation fields and high gas density.
In fact, diffuse warm neutral gas
($T\sim 8000$ K and $n\sim 1$ cm$^{-3}$;
McKee \& Ostriker 1977) occupies a
significant volume in the interstellar spaces
of galaxies and can be detected more easily
than cold gas (Hirashita et al.\ 2003).
Therefore, we also examine more diffuse gas
whose typical density is much smaller than
$100~{\rm cm}^{-3}$ in
Section \ref{subsec:warm}. Indeed,
the spin temperature of H {\sc i} in
Chengalur \& Kanekar (2000) is high
($T\ga 1000$ K) for a large part of their
sample, although the relation between the
spin temperature and the kinetic temperature
should be carefully discussed.

\section{DUST AND \H2}\label{sec:results}

\subsection{Various column densities}

We examine the relation between dust-to-gas ratio and
\H2 abundance of the sample Ly$\alpha$ systems. The
relation can be predicted by solving
equation (\ref{eq:equil}). First of all, we should
examine if the relation is compatible with the
reasonable physical state of gas. From the \H2
detected objects, we have derived
$n\sim 100~{\rm cm}^{-3}$, $T\sim 100$ K, and
$\chi\sim 10$. In Figure \ref{fig:h2_dust}, we show
the relation between $\fH2$ and $\kappa$ calculated
by our model with various H {\sc i} column densities
($n=100~{\rm cm}^{-3}$, $T=100$ K, and $\chi =10$
are assumed). The thick solid, dotted, dashed, and
dot-dashed lines represent the results
with $N_{\rm H}=10^{21}$, $10^{19}$, $10^{20}$, and
$10^{22}$ cm$^{-2}$, respectively. The thin solid
line represent the result with a high \H2 formation
rate on grains as suggested for some objects
(Section \ref{subsec:each}), where $a=0.03~\mu$m
is assumed to increase $R_{\rm dust}$ by 3.3 times
in order to see the effect of increased \H2
formation rate.

The relation between molecular fraction and dust-to-gas
ratio is well reproduced. The rapid enhancement of
molecules for $f_{\rm H_2}>\fH2^{\rm cr}$
(equation \ref{eq:critical}) is caused
by the self-shielding effect. If $N_{\rm H}$ is large,
the self-shielding condition is achieved with a small
value of $\fH2$. Therefore, the molecular fraction
tends to become larger in systems with larger
$N_{\rm H}$.

The molecular fraction is
very sensitive to the H {\sc i} column density
and the \H2 formation rate on dust grains in the
self-shielding regime. Because of such a sensitive
dependence, the large dispersion of $\fH2$
among the \H2 detected objects can be reproduced by
the four lines.

\subsection{Density and temperature}
\label{subsec:h2_dust_dep}

\begin{figure*}
\includegraphics[width=8cm]{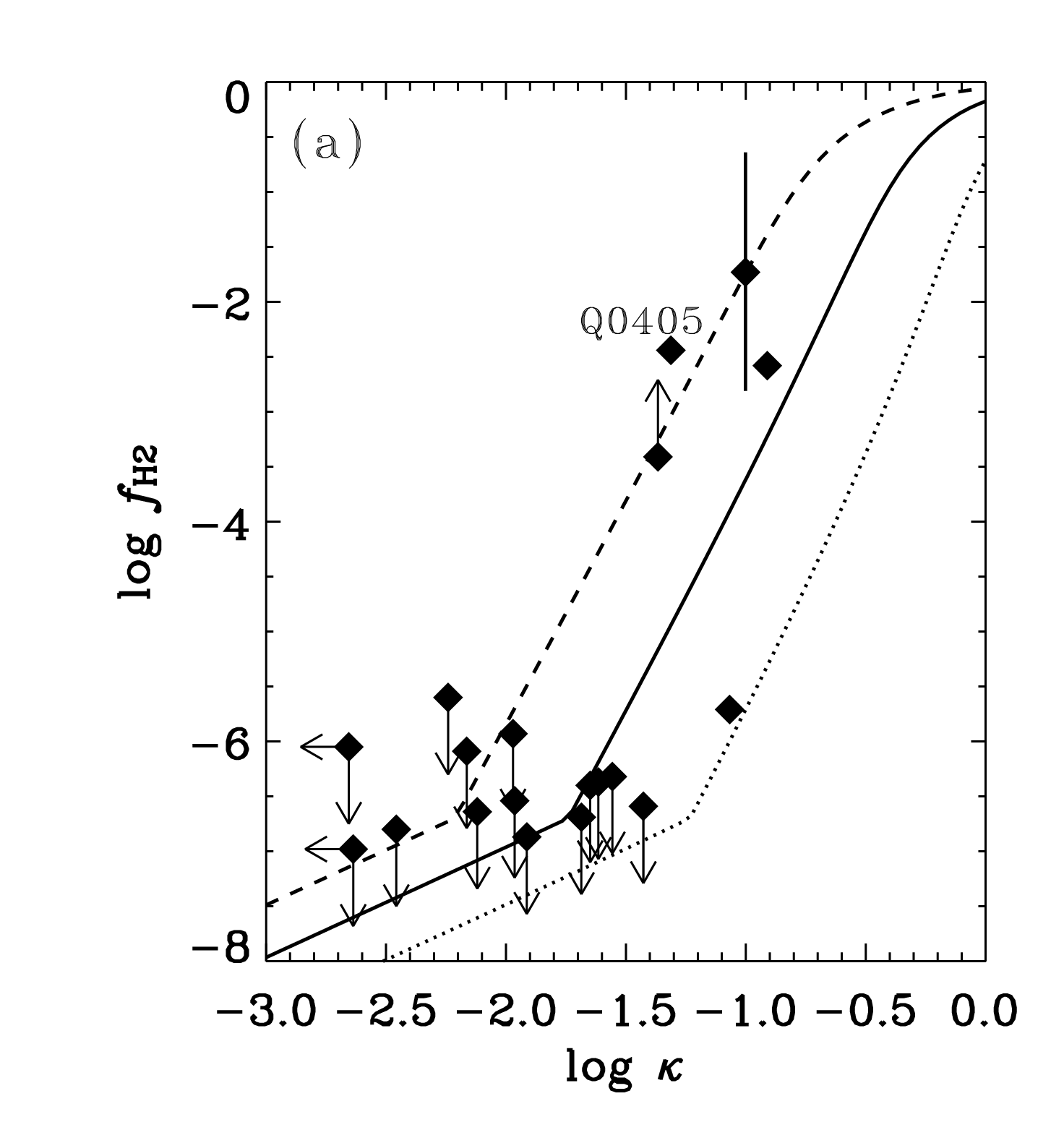}
\includegraphics[width=8cm]{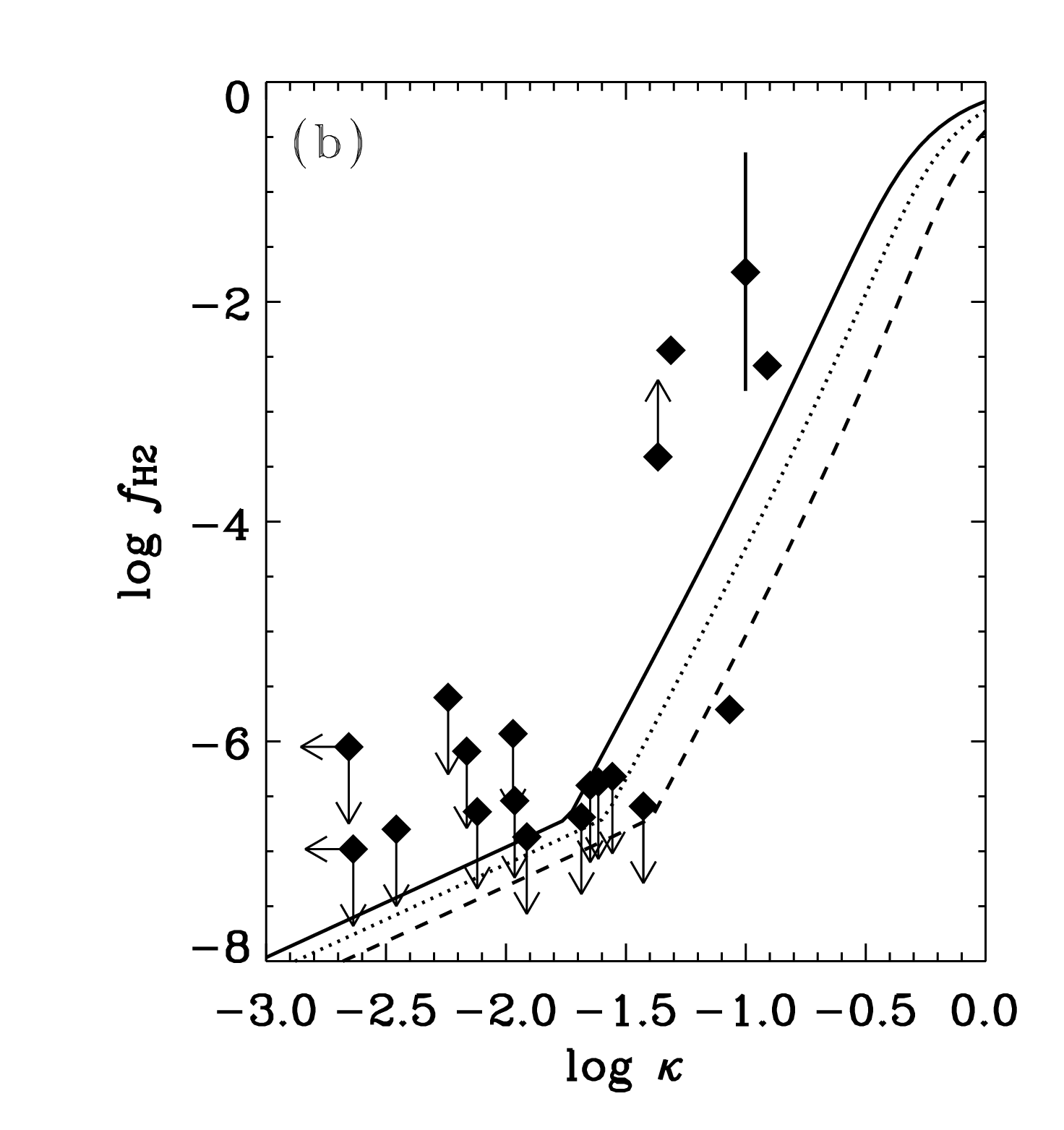}
\includegraphics[width=8cm]{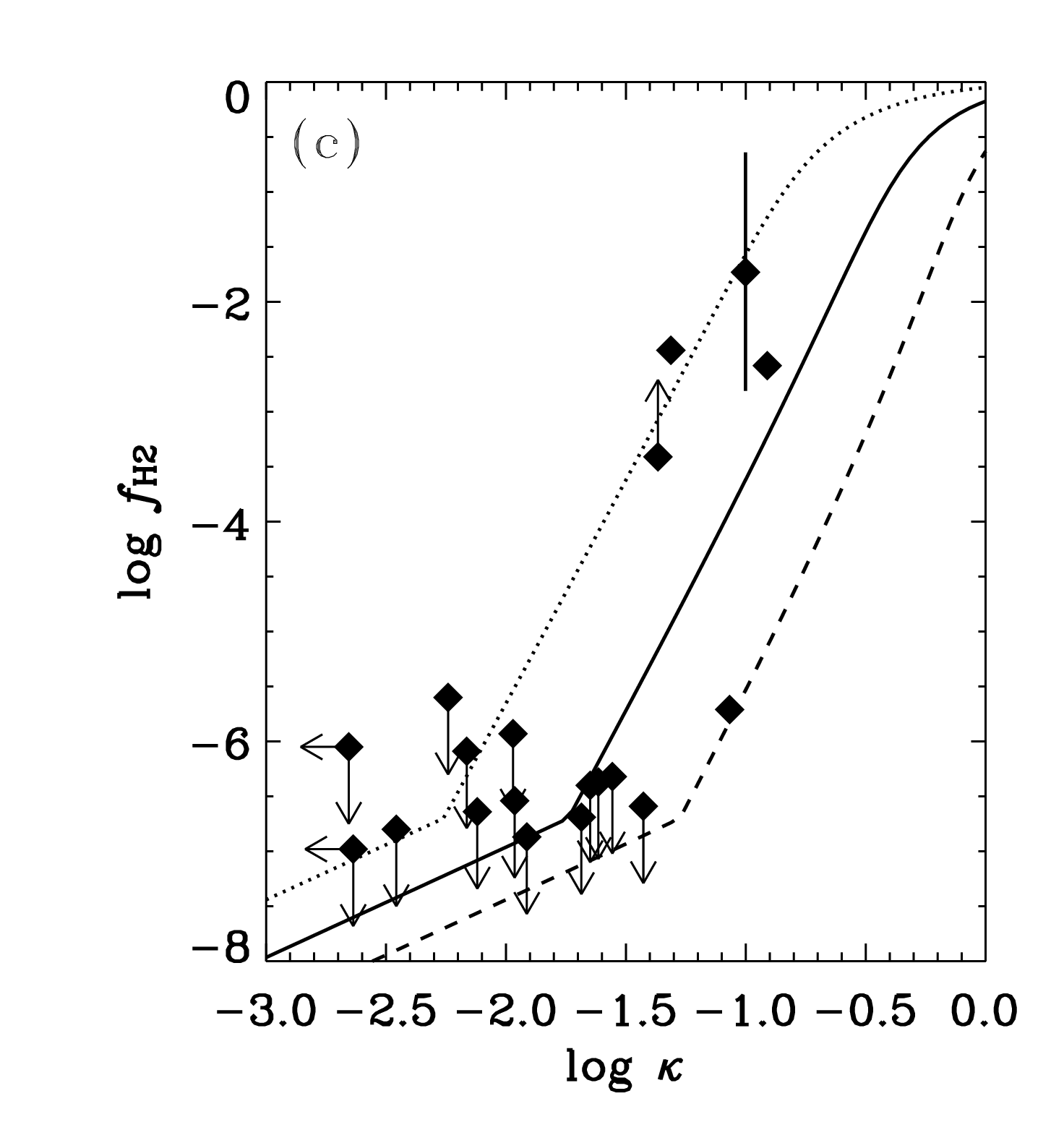}
\caption{Same relations as Figure \ref{fig:h2_dust}.
Only the data with $20.5<\log N_{\rm H}<21.5$ are
shown (squares). (a) The solid, dotted, and dashed
lines represent
our calculations for $n=100~{\rm cm}^{-3}$,
$n=30~{\rm cm}^{-3}$, and $n=300~{\rm cm}^{-3}$,
respectively. The other quantities are fixed:
$T=100$ K, $\chi =10$, and
$N_{\rm H}=10^{21}$ cm$^{-2}$. The DLA toward
Q 0405$-$443 ($z_{\rm abs}=2.595$), marked with
`Q0405', has an ISRF ($\chi <4.1$) much smaller
than assumed here. (b) The solid, dotted, and
dashed lines correspond to our results for
$T=100$ K, 30 K, and 1000 K, respectively, with
$n=100~{\rm cm}^{-3}$, $\chi =10$, and
$N_{\rm H}=10^{21}~{\rm cm}^{-3}$. (c) The solid,
dotted, and dashed lines represent our results for
$\chi =10$, 3, and 30, respectively, for
$n=100~{\rm cm}^{-3}$, $T=100$ K, and
$N_{\rm H}=10^{21}~{\rm cm}^{-2}$, respectively.
\label{fig:h2_dust_dep}}
\end{figure*}

Here, we examine the dependence on  gas density,
temperature, and ISRF. In the previous subsection,
we have shown that $\fH2$ depends on $N_{\rm H}$.
Therefore, we only concentrate on the objects with
$20.5<\log N_{\rm H}<21.5$. This range is typical
of DLAs.
In Figure \ref{fig:h2_dust_dep}, we show the
relation between $\fH2$ and
$\kappa$ for various (a) gas density, (b) gas
temperature, and (c) intensity of ISRF.

In Figure \ref{fig:h2_dust_dep}a, we investigate the
three densities $n=100~{\rm cm}^{-3}$, 30 cm$^{-3}$,
and 300 ${\rm cm}^{-3}$, in order to test if the
observational data points are reproduced with an
order-of-magnitude density range centered at the
typical density derived for \H2 detected objects
($n\sim 100~{\rm cm}^{-3}$). We assume typical
quantities: $T=100$ K, $\chi =10$, and
$\log N_{\rm H}=21$, and we adopt $a=0.1~\mu$m
and $\delta =2$ g cm$^{-3}$ (unless otherwise stated,
we adopt those $a$ and $\delta$ throughout this
paper). The data are well reproduced except for a
point marked with `Q0405'. This represents the DLA at
$z_{\rm abs}=2.595$ toward the quasar
Q 0405$-$443 ($\log N_{\rm H}=20.90$). However, for
this object, the
ISRF is estimated to be $\chi <4.1$
(Section \ref{subsec:each}), smaller than
assumed here ($\chi =10$). This low radiation field
is a possible reasons for the molecular abundance
larger than that predicted by the models, although
there could be other possibilities (e.g., large
$R_{\rm dust}$).

Figure \ref{fig:h2_dust_dep}b shows the dependence
on gas temperature. As discussed in
Section \ref{subsec:chemi}, the temperature dependence
of the \H2 formation rate $R_{\rm dust}$ is
uncertain. For example, if we take into account the
recombination efficiency
in Cazaux \& Tielens (2002), the \H2 formation rate is
much reduced for $T\ga 100$ K, and $\fH2$ becomes
more than 4 times smaller than presented in this
paper. Therefore, Figure \ref{fig:h2_dust_dep}b
is shown only to examine the conventional reaction
rate often assumed in other literatures.
The DLA toward Q 0347$-$383
at $z_{\rm abs}=3.025$ has the highest
$T_{10}\sim 800$ K of all the \H2 detected DLAs,
and we examine the temperature range from 30 K up to
1000 K. The solid, dotted, and dashed
lines in Figure \ref{fig:h2_dust_dep}b
correspond to $T=100$ K, 30 K, and 1000 K,
respectively.

The temperature may systematically
change as a function of dust-to-gas ratio
($\kappa$), because the photoelectric heating of
dust dominates the gas heating
(Wolfire et al.\ 1995). However, at least for
the \H2 detected objects, there is no evidence
that the gas temperature correlates with the
dust-to-gas ratio. In this paper, we do not
include such a correlation in our analysis.

Figure \ref{fig:h2_dust_dep}c shows the dependence on
the ISRF intensity. We examine $\chi =3$, 10, and 30,
in order to examine an order of magnitude centered at
the typical value derived for \H2 detected objects.
We see that the range well reproduces the observed data
points. The increase of $n$ has almost the same
influence on the decrease of $\chi$, i.e.,
the result become the same if we assume the same
$R_{\rm dust}n/\chi$. Indeed, the ratio between the
\H2 formation
and destruction rates is proportional to
$R_{\rm dust}n/\chi$.

The above results generally show that $\fH2$ is
sensitive to the variation of physical quantities
particularly in the self-shielding regime. The
sensitive dependence of $f_{\rm H_2}$ in the
self-shielding regime causes a large dispersion of
$f_{\rm H_2}$, and almost all the data points with a
large scatter are explained by the density and
temperature range considered above
(see also the discussion in
Section \ref{subsec:scaling}). The scatter
typically arises for $\log\kappa\ga -1.5$.

As a summary of this section, we present the
likelihood contours on the $\fH2 -\kappa$ diagram
under the condition that $n$, $T$, and $\chi$ vary
in the above range:
$1.5\leq\log n~[{\rm cm}^{-3}]\leq 2.5$,
$1.5\leq\log T~[{\rm K}]\leq 3$,
and $0.5\leq\log\chi\leq 1.5$.
Here the likelihood is defined as the number
of combinations of
$(\log n,\,\log T,\,\log\chi)$ that satisfies
a certain $(\log\fH2 ,\,\log\kappa)$. We follow
the formulation described in
Appendix \ref{app:likelihood}, where
we put $\bm{x}=(\log\fH2 ,\,\log\kappa)$,
$\bm{y}=(\log n,\,\log T,\,\log\chi)$,
$N=64$, and $M=64$ with the range of
$(\log n,\,\log T,\,\log\chi)$ described above
([1.5, 2.5], [1.5, 3], and [0.5, 1.5],
respectively) (the result is independent on
$N$ and $M$ if we adopt numbers larger than
$\sim 30$). As before, the observational
sample is limited to the DLAs with
$20.5<\log N_{\rm H}~[{\rm cm}^{-2}]<21.5$
are examined, and we assume
$N_{\rm H}=10^{21}~{\rm cm}^{-2}$ in the
theoretical calculation. For a more detailed
analysis, the probability distribution functions
of $(\log n,\,\log T,\,\log\chi)$ should be
considered. Since the probability distribution
functions are unknown for those quantities,
we only count the number of solutions. The
ranges constrained here could be regarded as the
typical dispersions ($\sigma$). The possible
physical correlation between those three
quantities is also neglected in our analysis.
We leave the modeling of the physical relation of
those quantities for a future work (see
Wolfire et al.\ 1995 for a possible way of
modeling).

\begin{figure}
\includegraphics[width=8cm]{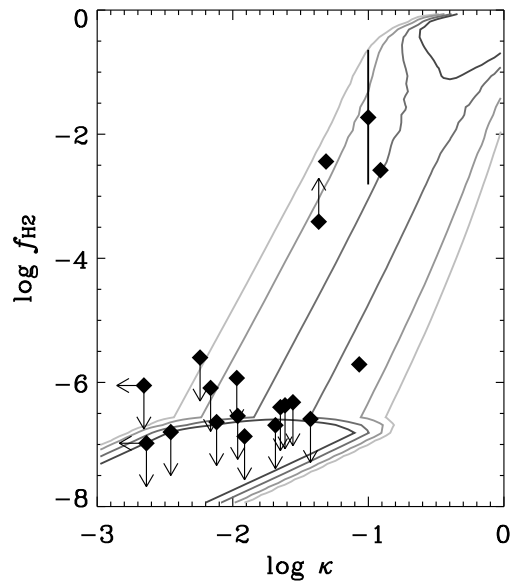}
\caption{Likelihood contour on the $\log\fH2 -\log\kappa$
diagram, when the density, temperature, and radiation field
are varied in the range
$1.5\leq\log n~[{\rm cm}^{-3}]\leq 2.5$,
$1.5\leq\log T~[{\rm K}]\leq 3$, and
$0.5\leq\log\chi\leq 1.5$, respectively. We adopt
$N_{\rm H}=10^{21}$ cm$^{-2}$. The contours
are drawn for four levels:
95\%, 90\%, 70\%, and 50\% of the sample is supposed to
be in the regions corresponding to each level. The
observational data points are the same as those in
Figure \ref{fig:h2_dust_dep}.
\label{fig:likeli}}
\end{figure}

In Figure \ref{fig:likeli}, we show the contour
of the likelihood $P$
(Appendix \ref{app:likelihood}). The levels show
likelihood contours of 
50\%, 70\%, 90\%, and 95\% according to  
our model (see Appendix \ref{app:likelihood}).
All the data points are explained by the assumed
ranges of the quantities. The wide range of
$\fH2$ covered in the self-shielding regime
($\log\fH2 >-6.7$) explains the observational
large scatter of $\fH2$.

\subsection{Column density}\label{subsec:column}

\begin{figure*}
\includegraphics[width=12cm]{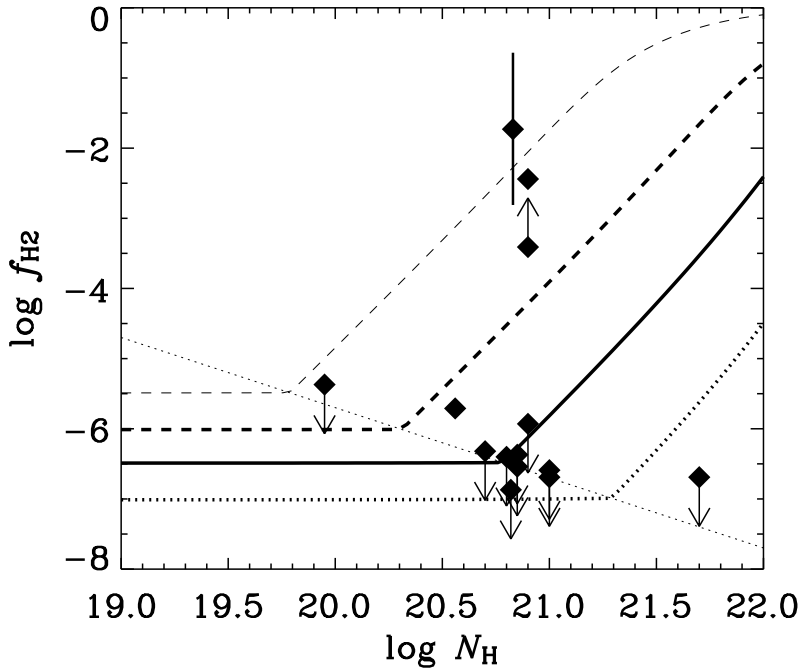}
\caption{Relation between the molecular fraction and
the H {\sc i} column density. Only the observational
sample with $0.01<\kappa <0.1$ is selected
(squares). The upper and lower arrows represents
the observational lower and upper limits for
the molecular fraction, respectively. The thick solid,
dotted, and dashed lines present our
calculations with various gas density
($n=100$, 30, and 300, respectively). The gas
temperature, dust-to-gas ratio and ISRF intensity are
assumed to be $T=100$ K, $\kappa =0.03$, and $\chi =10$,
respectively. The thin dashed line indicate the same
calculation with the thick dashed line, but a higher
dust-to-gas ratio of $\kappa =0.1$ is assumed.
We also show the self-shielding condition by dotted
line, above which the \H2 self-shielding effect
becomes significant.
\label{fig:h2_N_various}}
\end{figure*}

Another important conclusion derived by
Ledoux et al.\ (2003) is that the H {\sc i}
column density and the molecular fraction do not
correlate. Therefore, we also present our model
calculation for the $N_{\rm H}-\fH2$ relation.
Since the
difference in the dust-to-gas ratio reproduces a very
different result, we only use the sample with
$0.01<\kappa <0.1$. Here we assume $\kappa =0.03$.
In Figure \ref{fig:h2_N_various}, we show our results
with various gas density
(the thick solid, dotted, and dashed lines represent the
results with $n=100$, 30, and 300 cm$^{-3}$,
respectively), where we assume that $T=100$ K,
$\kappa =0.03$, and $\chi=10$. The thick lines
underproduces the observed \H2 fraction of the
\H2 detected objects, since the dust-to-gas ratio of
those objects are systematically higher than that
assumed. In particular, the data point with the bar
($z_{\rm abs}=1.973$ toward Q 0013$-$004) has the
dust-to-gas ratio of $\kappa =0.099$. This data point
can be reproduced by the thin dashed line produced with
the same condition as the dashed line but with
$\kappa =0.1$. As seen in
Section \ref{subsec:h2_dust_dep}, the same
$n/\chi$ reproduces the same result with the other
quantities fixed. Thus, we do not
show the result for the various ISRF intensity $\chi$.

The thin dotted line in Figure \ref{fig:h2_N_various}
represents the relation $\fH2 =\fH2^{\rm cr}$
(equation \ref{eq:critical}). Therefore, if a data
point is above the thin dotted line, the \H2 is
self-shielding the dissociating photons.
We observe that the
molecular fraction is very sensitive to the gas
density and ISRF intensity,
especially in the self-shielding regime. This
sensitive dependence tends to erase the correlation
between $f_{\rm H_2}$ and $N_{\rm H}$ in the
observational sample and could explain the
absence of correlation in the observational
sample.

\subsection{Possibility of warm phase}\label{subsec:warm}

As mentioned in Section \ref{subsec:typical}, our derived
quantities are biased to the \H2 detected objects. As shown
in Hirashita et al.\ (2003), \H2-rich regions
are only confined in a small dense regions, whose
gas temperature is $T\la 100$ K. However,
Chengalur \& Kanekar (2000) observationally derive the
spin temperature $T\ga 1000$ K for
a large part of their DLA sample and $T\sim 100$ K for a few
DLAs, although the large beam size relative to the size of
the QSO may tend to overestimate the spin temperature.
Ledoux et al.\ (2002) find that the DLAs with \H2
detection are always dense ($n>30~{\rm cm}^{-3}$) and cool
($T<100$ K). Therefore, most of the DLAs may be warm
except for the \H2 detected ones.

Based on a simulation suitable for DLAs,
Hirashita et al.\ (2003) have shown that most ($\sim 90$\%)
of the regions are covered by warm and diffuse regions with
$T\sim 1000$--10000 K and $n\sim 1$ cm$^{-3}$. In the warm
phase, \H2 formation on dust is not efficient
and \H2 formation occurs in gas phase (Liszt 2002;
Cazaux \& Tielens 2002). Therefore, we cannot put any
constraint on the physical state of
warm gas in the framework of this paper.
The \H2 formation in gas phase occurs in the following
route: ${\rm H}+{\rm e}^-\to{\rm H}^-+h\nu$,
${\rm H}^-+{\rm H}\to{\rm H}_2+{\rm H}^+$,
and ${\rm H}^++{\rm H}\to{\rm H}_2^+$,
${\rm H}_2^++{\rm H}\to{\rm H}_2+{\rm H}^+$.
Liszt (2002) shows that the gas phase reactions
result in a molecular fraction
$\fH2\sim 10^{-7}$--$10^{-8}$. This range is
consistent with the data with upper limits of $\fH2$.

\subsection{Lack of very \H2-rich DLAs?}

The likelihood contours presented in
Figure \ref{fig:likeli} suggests that some DLAs
should be rich in \H2 ($\log\fH2\ga -1$) if the
dust-to-gas ratio is around the Galactic value
($\log\kappa\ga -0.5$). However, all the objects in
the sample have molecular fraction
$\log\fH2 <-1$. There are three possible explanations that
we discuss in the following. 

The first possibility of the lack of very 
molecule-rich DLAs is the contamination from the
molecule-poor intercloud medium. If the
contribution of
intercloud medium to the column density is high,
the molecular fraction is inferred  to be low
even if a molecule-rich region is present along the
line of sight.

Secondly, a QSO selection effect might occur. If the
dust-to-gas ratio is as high as $\log\kappa >-0.5$
and the column density is
$N_{\rm H}=10^{21}~{\rm cm}^{-2}$, the optical
depth of dust in UV is larger than 0.3
(equation \ref{eq:tauUV}). Therefore, QSO is
effectively extinguished by dust if there is
a dust-rich cloud in the line of sight.
Such a population is also suggested by a
numerical simulation (Cen et al.\ 2003).
Observationally, it is a matter of debate
whether the dust bias is large or not.
Ellison et al.\ (2002)
study optical colours of
optically-selected QSO samples, and find that
the effect of dust extinction of intervening
absorbers is small. Fall et al.\ (1989) show
a significant dust extinction of intervening DLAs
by showing that QSOs with foreground DLAs tend to
be redder than those without foreground DLAs.

The third possibility is concerned with the
probability of detecting molecule-rich DLAs.
Hirashita et al.\ (2003) have shown that the
covering fraction of \H2-rich clouds in a galactic
surface is $\la 10$\%.
Therefore, the probability
that the line of sight passes through such clouds
may be very low. Some
very small molecule-rich clumps, which would be
difficult to find as QSO absorption systems, are also
found (e.g.\ Richter et al.\ 2003; Heithausen 2004).
The probability distribution
function of gas density and temperature should
be treated by taking into account the covering
fraction. The detailed treatment of such probability
is left for the future work.

Extremely molecular-rich objects with $\fH2\sim 1$
might escape H {\sc i} absorption detection
(Schaye 2001). A new strategy will be required to
detect such fully molecular clouds at high $z$.
An observational strategy for
high-$z$ molecular clouds with high column densities
is discussed in Shibai et al.\ (2001).

\subsection{Summary of our analysis}

\begin{figure}
\includegraphics[width=8cm]{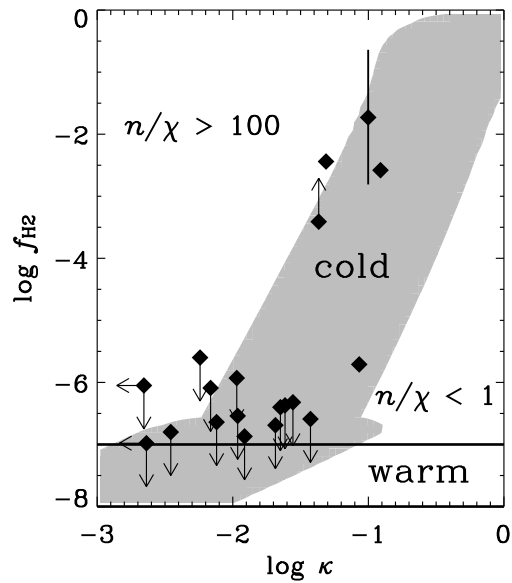}
\caption{Summary of our analysis.
If we obtain the molecular fraction and dust-to-gas
ratio for an absorption system, we can roughly
estimate the physical state of gas. This diagram
should only be used to obtain the first rough estimate.
The region marked with ``cold'' indicates the same
area as the 90\% level of Figure \ref{fig:likeli}
(representative of a typical cold phase). The strip
marked with ``warm'' indicates the
typical range of $\fH2$ in a warm phase where \H2 is
formed in gas phase (Liszt 2002). The other two
regions are possibly characterised by the physical
state, $n/\chi\ga 100$ and $n/\chi\la 1$ (see text
for details).
\label{fig:summary}}
\end{figure}

In order to summarise our analysis and for the
observers' convenience, we present
Figure \ref{fig:summary}. Various physical states
of gas could be discriminated on the $\fH2 -\kappa$
(molecular fraction vs.\ dust-to-gas ratio) diagram.
First of all, we should stress that this diagram is
only useful to obtain the first result about the
gas state of an absorption line system whose \H2
fraction and dust-to-gas ratio are constrained,
or to obtain the statistical properties of gas
phase of a sample of absorption systems. For
the confirmation of gas state, more detailed
analysis such as treatment of C {\sc i} fine
structure lines should be combined.

In Figure \ref{fig:summary}, the shaded area
shows the region where 90\% of the gas
with $1.5\leq\log n~[{\rm cm}^{-3}]\leq 2.5$,
$1.5\leq\log T~[{\rm K}]\leq 3$, and
$0.5\leq\log\chi\leq 1.5$ is predicted to be
located (see Figure \ref{fig:likeli}). Those
ranges of the quantities are typical of
\H2-detected objects and representative of
``cold'' gas. The strip indicated by ``warm''
shows the warm phase in which \H2 predominantly
forms in
the gas phase (we take the values from Liszt 2002).
There is an overlapping region of the cold and
warm states, and if a data point lies in this
region, the warm and cold states are equally
probable. There remain two regions not included
neither in ``cold'' nor in ``warm''. The upper
region shows an enhancement of the molecular
fraction, which requires a high \H2 formation
rate and/or a low \H2 destruction rate.
This condition is typically characterised with
$n/\chi\ga 100$ if the gas temperature is
favourable for the \H2 formation on dust grains
($\la 100$ K).
On the other hand, the lower region marked with
``$n/\chi <1$'' indicates that the \H2 fraction
is larger than the typical value in warm phase
but lower than our likelihood range for cold gas.
If a data point lies in this region, there could
be the following three possibilities: (i)
$n/\chi\la 1$ and $T\la 100$ K, so that the
\H2 formation rate is suppressed because of
a low density and/or the \H2 destruction rate
is enhanced because of a high ISRF;
(ii) $300\la T\la 500$ K, so that the \H2
formation on dust occurs only with a small
rate; (iii) the cold and warm phases coexist
in the line of sight. We can use this diagram
as the ``first
quick look'' for the physical state of gas if
\H2 is detected in a system whose dust-to-gas
ratio has been estimated.

\section{STAR FORMATION}\label{sec:sfr}

\subsection{Star formation rate}\label{subsec:sfr}

The above results for the relation between molecular abundance
and dust-to-gas ratio for DLAs strongly suggest that there
are internal UV radiation sources originating from star formation
activity (see also Ge \& Bechtold 1997; Ledoux et al.\ 2002;
Petitjean et al.\ 2002). Indeed, the cosmic UV
background radiation intensity is typically
$J_{21}\sim 1$ around $z\sim 3$. This corresponds to
$\chi\simeq 3.1\times 10^{-2}$. The ISRF intensity is
clearly larger than this, and the local heating sources
such as stars dominates the ISRF. Assuming that the ISRF
is produced by stars, we relate the ISRF with the SFR.

Hirashita, Buat, \& Inoue (2003) have derived the relation
between UV luminosity and SFR as
\begin{eqnarray}
{\rm SFR}=C_{2000}L_{2000}\, ,
\end{eqnarray}
where $L_{2000}$ is the monochromatic luminosity at
2000 \AA\ and $C_{2000}=2.03\times 10^{-40}$
[($M_\odot$ yr$^{-1}$)/(erg s$^{-1}$ \AA$^{-1}$)]
(under a Salpeter initial mass function with the stellar
mass range of 0.1--100 $M_\odot$ and a constant star
formation rate with the duration of $10^8$ yr;
see also Iglesias-P\'{a}ramo et al.\ 2004).
The surface luminosity density, defined as the luminosity
per surface area, can be roughly equated with the
ISRF intensity (Appendix \ref{app:surface}). Therefore,
the 2000 \AA\ surface monochromatic luminosity density,
$\Sigma_{\rm 2000}$, is estimated as
\begin{eqnarray}
\Sigma_{2000}\simeq cu_{2000}=9.0\times 10^{-7}\chi~
{\rm erg~cm}^{-2}~{\rm s}^{-1}~\mbox{\AA}^{-1}\, .
\label{eq:surfaceUV}
\end{eqnarray}
where we use the 2200 \AA\ energy density in Habing (1968)
for the normalisation of $u_{2000}$ (2000 \AA\
monochromatic radiative energy density).
Using equation (\ref{eq:surfaceUV}), we obtain the surface
density of SFR, $\Sigma_{\rm SFR}$:
\begin{eqnarray}
\Sigma_{\rm SFR}=C_{2000}\Sigma_{2000}=1.7\times 10^{-3}
\chi~M_\odot~{\rm yr}^{-1}~{\rm kpc}^{-2}\, .
\label{eq:sigma_SFR}
\end{eqnarray}
Indeed, this roughly gives the Galactic surface SFR
density
($\sim 10^{-3}~M_\odot~{\rm yr}^{-1}~{\rm kpc}^{-2}$;
Burkert, Truran, \& Hensler 1992)
if we assume a typical ISRF intensity of the solar
vicinity ($\chi =1$). But some observational data indicate
higher Galactic surface SFR in the solar vicinity such as
$\sim 5\times 10^{-3}~M_\odot~{\rm yr}^{-1}~{\rm kpc}^{-2}$
(Smith et al.\ 1978).
The general ISRF in the Galaxy could be systematically
higher (i.e.\ $\chi > 1$; Shibai et al.\ 1999).

Wolfe et al.\ (2003) have derived the calibration
$\Sigma_{\rm SFR}=2.5\times 10^{-3}\chi~M_\odot~
{\rm yr}^{-1}~{\rm kpc}^{-2}$. The difference comes
from the different assumption on the IMF and the
different stellar mass-luminosity relation. However
the difference in only 0.17 in the logarithmic scale.

The probable range of the radiation field constrained
in section \ref{subsec:h2_dust_dep} is
$0.5\la\log\chi\la 1.5$. This range predicts the
surface SFR density,
$5\times 10^{-3}~M_\odot~{\rm yr}^{-1}~{\rm kpc}^{-2}
\la\Sigma_{\rm SFR}\la 5\times
10^{-2}~M_\odot~{\rm yr}^{-1}~{\rm kpc}^{-2}$.
Wolfe et al.\ (2003a) derive the UV radiation field from
C {\sc ii}* absorption line. Their analysis is dependent
on the assumed
phase (cool or warm) of the ISM. The calculated SFR
differs by an order of magnitude between the cool and
warm media. Based on their results,
Wolfe et al.\ (2003b) have suggested that the probable
range of the SFRs of their sample
is $10^{-3}\la\Sigma_{\rm SFR}~
[M_\odot~{\rm yr}^{-1}~{\rm kpc}^{-2}]\la 10^{-2}$. This
range is roughly consistent with our range considering
the uncertainty in the assumed physical state of gas.
A numerical work by
Nagamine, Springel, \& Hernquist (2004) explains the
SFR theoretically.
Our estimate provides an independent observational
calculation for the SFR of DLAs.

Assuming a typical radius of $R=3$ kpc for DLAs
(Kulkarni et al.\ 2000), we obtain the SFR
$\Sigma_{\rm SFR}\pi R^2\sim 0.1$--1
$M_\odot~{\rm yr}^{-1}$. This range is broadly
consistent with the
upper limits obtained by some imaging observations
of DLAs (Bunker et al.\ 1999; Kulkarni et al.\ 2000;
Bouch\'{e} et al.\ 2001).

\subsection{Comparison with other SFR estimates}

Wolfe et al.\ (2003a) have estimated the SFR of a
sample of DLAs by using C {\sc ii}* absorption
intensity. They
consider the balance between the cooling rate dominated
by [C {\sc ii}] fine structure line emission and the
heating rate dominated by photoelectric
heating of a UV radiation field. Since our method for
SFR estimate is independent of theirs, the consistency
between the two methods is interesting to
explore. Wolfe et al.\ (2003a) investigate two thermally
stable states, WNM (warm neutral medium) and CNM
(cold neutral medium), adopting the scheme of
Wolfire et al.\ (1995).
There are three overlapping samples between
Wolfe et al.\ (2003a) and Ledoux et al.\ (2003) as shown
below.

\subsubsection{Q\,0347$-$383 ($z_{\rm abs}=3.025$)}

Some \H2 lines are detected in this object, and we
have derived $\chi =15$ (Section \ref{subsec:each}),
which is equivalent to
$\log\Sigma_{\rm SFR}~[M_\odot~{\rm yr}^{-1}~
{\rm kpc}^{-2}]=-1.6$. Wolfe et al.\ (2003a) derive
$\log\Sigma_{\rm SFR}~[M_\odot~{\rm yr}^{-1}~
{\rm kpc}^{-2}]=-1.3$ and $-2.2$ for the WNM and CNM,
respectively. Our estimated temperature
$T=400$--800 K is higher than the typical value for
CNM ($\sim 100$ K) and lower than the WNM.
In any case, our estimate is between the two values
of Wolfe et al.\ (2003a).3

\subsubsection{Q\,1223+178 ($z_{\rm abs}=2.466$)}

For this object, the depletion factor is extremely
small (${\rm [Fe/Zn]}=-0.07\pm 0.21$), and the
metallicity is also small
(${\rm [Zn/H]}=-1.63\pm 0.11$)
(Ledoux et al.\ 2003). Thus, it is expected that
dust-to-gas ratio of this object is extremely small.
Moreover, if we assume the above metal abundances, we
obtain $\log\kappa =-2.46$. With this dust abundance,
the  formation of \H2 through H$^-$ could be an
important process. The SFR cannot be estimated by our
method.

\subsubsection{Q\,1232$+$082 ($z_{\rm abs}=2.338$)}

In Section \ref{subsec:each}, we have estimated the
radiation field to be $\chi\leq 64$, which is
equivalent to the surface SFR density
$\log\Sigma_{\rm SFR}~[M_\odot~{\rm yr}^{-1}~
{\rm kpc}^{-2}]\leq -0.96$. The low excitation
temperature ($T_{10}=185$ K) indicates that the gas
in the CNM. The CNM solution of Wolfe et al.\ (2003a)
shows
$\log\Sigma_{\rm SFR}~[M_\odot~{\rm yr}^{-1}~
{\rm kpc}^{-2}]=-1.9$, consistent with our upper
limit.

\subsection{Cosmological implications}

Wolfe et al.\ (2003b) have extended their discussion on
the SFR in DLAs to a cosmological context. They have
found that the hypothesis that most of the DLAs
are in WNM is ruled out because it would conflict with
background light constraint. On the other hand,
Chengalur \& Kanekar (2000) have observationally
shown that a large part of their DLAs have a temperature
similar to WNM. Hirashita et al.\ (2003), by using detailed
numerical simulations, have also argued that the
probability to observe the CNM of a DLA
is small because the tiny covering fraction of such phase.
Their calculation also shows that \H2 detected objects
are biased to the CNM, and is consistent with the
estimates in Section \ref{subsec:each}. Most of the
\H2 deficient DLAs may be in the WNM.
Thus at the moment, there seems to be tension between the 
CNM and WNM hypotheses, which requires additional work in order
to be relaxed. A composite analysis of fine-structure excitation
and \H2 may be required to confirm our results of a simple
statistical approach.

In spite of such an uncertainty, it is interesting to note
that the star formation activity in DLAs can be investigated
via our treatment of \H2 formation and destruction rates.
The surface SFR density derived by us is comparable to that
inferred in
Wolfe et al.\ (2003a), who have suggested
that DLAs are an important population in the cosmic
star formation (and metal production) history (see also
Pei \& Fall 1995;
Pei, Fall, \& Hauser 1999). The SFR is
much lower than forming elliptical galaxies as calculated
by Arimoto \& Yoshii (1987), but more similar to
spiral galaxies. In the context of the hierarchical
structure formation,
Nagamine et al.\ (2004)
have explained the SFR of DLAs by a cosmological simulation,
and they also show that the
SFR of a DLA is generally smaller than that of a
typical Lyman break galaxies
($\sim 50~M_\odot~{\rm yr}^{-1}$) (but see Schaye 2004).
According to their
simulation, the host halo mass of DLAs spans over a
wide range from $\sim 10^9~M_\odot$ to
$\sim 10^{12}~M_\odot$. Another theoretical calculation
by Hirashita \& Ferrara (2002) show that high-$z$
dwarf galaxies whose total mass of dark halo is around
$10^9~M_\odot$ has a SFR similar to
that estimated in Section \ref{subsec:sfr}.

Our statistical work in this paper will be extended to the
cosmological star formation history and some observational
consequences (see also Ferrara et al.\ 1999).
The redshift evolution of \H2 abundance and the relation
between metals and \H2 (Curran et al.\ 2004) can also be
used to constrain the cosmological SFR.

\section{SUMMARY}\label{sec:sum}

We have modeled the \H2 abundance of QSO absorption
line systems and compared our model calculations with
observational samples of damped Ly$\alpha$ systems
(DLAs) and sub-DLAs. We have derived the \H2 abundance,
$ f_{\rm H_2}$, as a function of dust-to-gas ratio,
$\kappa$ (normalized to the Galactic value) considering
\H2 self-shielding and dust extinction against
dissociating photons. The $\fH2 -\kappa$ relation
depends on the gas density ($n$) and temperature ($T$),
and the ISRF intensity
($\chi$: normalised to the Galactic value). Our aim
has been to constrain those quantities by using \H2
data. Treating the data of \H2 excitation states of
the \H2 detected objects, we adopt
$1.5\la\log n~[{\rm cm}^{-3}]\la 2.5$,
$1.5\la\log T~[{\rm K}]\la 3$, and
$0.5\la\log\chi\la 1.5$. From the
comparison with data, we have found that the observational
$f_{\rm H_2}$ -- $\kappa$ relation is naturally explained
by the above range. The efficient photodissociation by
the ISRF can explain
the extremely small \H2 abundance ($\fH2\la 10^{-6}$)
observed for $\kappa\la 1/30$. We have also succeeded in
explaining the rapid increase of \H2 abundance for
$\kappa \ga 1/30$
by the effect of self-shielding of \H2 dissociating photons.
Because of a nonlinear dependence of self-shielding on
the physical quantities, a large scatter of \H2 fraction is
reproduced. However, we
should note that the above parameter range
may be biased to the cold gas favourable for \H2 formation.
It is still possible that most of the \H2 deficient
DLAs and sub-DLAs might be in a diffuse and warm state.

We finally propose to estimate star formation rates (SFRs)
of (sub-)DLAs from \H2 observations. The SFRs estimated by
our method are compared with those derived by
Wolfe et al.\ (2003a). Two common samples give roughly
consistent SFRs. The strength of UV field indicates a
surface SFR density:
$5\times 10^{-3}$--$5\times 10^{-2}~M_\odot~{\rm yr}^{-1}
~{\rm kpc}^{-2}$. Therefore, DLAs are actually star-forming
objects. The SFR is smaller than typical Lyman break
galaxies, but DLAs may be a major population responsible
for star formation in the high-$z$ universe.

\section*{Acknowledgments}
We thank the anonymous referee for helpful comments and
I. T. Iliev, J. X. Prochaska, P. Richter, H. Shibai, 
P. Petitjean, and A. Wolfe for useful discussions. HH is supported by
JSPS Postdoctoral Fellowship. We fully utilised the
NASA's Astrophysics Data System Abstract Service (ADS).

\appendix

\section{SURFACE LUMINOSITY DENSITY AND RADIATION FIELD}
\label{app:surface}

In the text, we have equated the surface luminosity
density $\Sigma$ with the interstellar radiation field (ISRF)
intensity $cu$.
However, it is not obvious a priori that this assumption is correct.
If the radiation from a stars is isotropic and the dust extinction
is neglected, the ISRF intensity at the
position $\bmr$, $cu(\bmr)$, is expressed as
\begin{eqnarray}
cu(\bmr)=\int d^3\bmr '
\frac{\rho (\bmr ')}{4\pi |\bmr ' -\bmr |^2}
\, ,
\end{eqnarray}
where $\rho (\bmr )$ is the spatial luminosity density of stars
($\rho(\bmr )d^3\bmr$ is the luminosity in the volume
element $d^3\bmr$).

One of the largest uncertainties is the spatial distribution of
radiating sources. Therefore,
in the following discussions, we derive the relation
between  $\Sigma$ and $cu$ under two representative
geometries of source distribution: spherical and
disc-like distributions.

\subsection{Spherical distribution}

In the spherical symmetric distribution, we can calculate the
ISRF intensity at the centre of the sphere by
\begin{eqnarray}
cu=\int_0^{R_0}dR\, 4\pi R^2\frac{\rho (R)}{4\pi R^2}=
\int_0^{R_0}dR\,\rho (R)\, ,
\end{eqnarray}
where $R_0$ is the radius of the whole emitting region.
The right-hand side gives a rough estimate of the surface
luminosity density (denoted as $\Sigma$), and therefore,
the ISRF can be equated with the surface luminosity
density (i.e.\ $\Sigma\simeq cu$).

\subsection{Disc-like distribution}

In the disc-like distribution, we can calculate the
ISRF intensity at the centre of the disc by
\begin{eqnarray}
cu=\int_0^{R_0}dR\, 2\pi R\int_{-h}^{h}dz
\frac{\rho (R,\, z)}{4\pi (R^2+z^2)}\, ,
\end{eqnarray}
where $2h$ is the disc thickness, and $R_0$ is the disc
radius. If we assume for simplicity that $\rho$ is
constant, we obtain analytically
\begin{eqnarray}
cu=\rho h\left\{\frac{R_0}{h}\arctan\frac{h}{R_0}+
\frac{1}{2}\ln\left[\left(\frac{R_0}{h}\right)^2+1\right]
\right\}\, .
\end{eqnarray}
The galactic discs usually satisfy $R_0\gg h$, so that
the above equation is approximated as
\begin{eqnarray}
cu\simeq \rho h\left(1+\ln\frac{R_0}{h}\right)\, .
\end{eqnarray}
The dependence on $R_0/h$ is very weak, and even if
we assume a very thin disc such as $R_0/h=100$, we obtain
$cu\simeq 5.6\rho h$.
The surface luminosity density can be typically estimated
by $\Sigma\simeq 2\rho h$. Therefore, $cu$ and $\Sigma$
have the same order of magnitude (different by a factor
of $\la 3$).

The above discussions justify our simple assumption
$cu=\Sigma$ in the text.

\section{LIKELIHOOD FORMULATION}
\label{app:likelihood}

We consider a set of $n$ physical quantities,
$\bm{x}=(x_1,\,\cdots ,\, x_n)$.
Suppose that those quantities are determined by a
set of $m$ parameters,
$\bm{y}=(y_1,\,\cdots ,\, y_m)$, each of which
has a reasonable range
$y_i^{\rm min}\leq y_i\leq y_i^{\rm max}$
($i=1,\,\cdots ,\, m$). The two sets of quantities
are related to the following function $f$:
\begin{eqnarray}
\bm{y}=f(\bm{x})\, .
\end{eqnarray}

We divide $\bm{x}$ and
$\bm{y}$ into $N+1$ and $M+1$ bins, respectively:
\begin{eqnarray}
x_i^j & \equiv & x_i^{\rm min}+\frac{j}{N}
(x_i^{\rm max}-x_i^{\rm min})~~~
(j=0,\,\cdots ,\, N)\, . \\
y_i^k & \equiv & y_i^{\rm min}+\frac{k}{M}
(y_i^{\rm max}-y_i^{\rm min})~~~
(k=0,\,\cdots ,\, M)\, .
\end{eqnarray}
The $j$-th and $k$-th bins of $x_i$ and $y_i$
can be defined as the range
$[x_i^{j-1},\, x_i^j]$ ($j=1,\,\cdots ,\, N$) and
$[y_i^{k-1},\, y_i^k]$ ($k=1,\,\cdots ,\, M$),
respectively.
Any $N$-dimensional bin of $\bm{x}$ can be
specified by a set of $N$ integers
$(j_1,\,\cdots ,\, j_n)$. We represent the
value of $\bm{y}$ in each bin with the median
as
\begin{eqnarray}
\bm{y}(k_1,\,\cdots ,\, k_m)\equiv\left(
\frac{y_1^{k_1-1}+y_1^{k_1}}{2},\,\cdots ,\,
\frac{y_m^{k_m-1}+y_m^{k_m}}{2}\right)\, .
\end{eqnarray}
We define $N(j_1,\,\cdots ,\, j_n)$ as the number
of the combination of $m$ integers
$(k_1,\,\cdots ,\, k_m)$ such that
$f(\bm{y}(k_1,\,\cdots ,\, k_m))$ is in the
$n$-dimensional bin
$(j_1,\,\cdots ,\, j_n)$ of $\bm{x}$. Then the
likelihood
of $\bm{x}(j_1,\,\cdots ,\, j_n)$,
$P(j_1,\,\cdots ,\, j_n)$, can be defined as
\begin{eqnarray}
P(j_1,\,\cdots ,\, j_n)\equiv
\frac{N(j_1,\,\cdots ,\, j_n)}
{\sum_{j_1,\,\cdots ,\, j_n}N(j_1,\,\cdots ,\, j_n)}
\, .
\end{eqnarray}

Let $\partial S(p)$ be a contour surface of $P=p$
($(n-1)$-dimensional surface), and let $S(p)$ be
the area where surrounded by $\partial S(p)$.
Then, the following sum Prob($S$) gives the probability
that the data $\bm{x}$ lies in $S$:
\begin{eqnarray}
{\rm Prob}(S)\equiv\sum_{\bm{x}\in S}P(\bm{x}(j_1,\,
\cdots ,\, j_n))\, .
\end{eqnarray}


\begin{thebibliography}{99}
\bibitem[Abel et al.(2004)]{abel04} Abel, N. P., Brogan, C. L.,
    Ferland, G. J., O'Dell, C. R., Shaw, G., \&
    Troland, T. H. 2004, ApJ, 609, 247
\bibitem[Abel et al.(1997)]{abel97} Abel, T., Anninos, P., Zhang, Y.,
    Norman, M. L. 1997, NewA, 2, 181
\bibitem[Andr\'{e} et al.(2004)]{andre04} Andr\'{e}, M. K. et al.\
    2004, A\&A, 422, 483
\bibitem[Arimoto \& Yoshii(1987)]{arimoto87} Arimoto, N., \&
    Yoshii, Y. 1987, A\&A, 173, 23
\bibitem[Bianchi et al.(2001)]{bianchi01} Bianchi, S., Cristiani, S.,
    \& Kim, T.-S. 2001, A\&A, 376, 1
\bibitem[Black et al.(1987)]{black87} Black, J. H., Chaffee,
    F. H. Jr., \& Foltz, C. B. 1987, ApJ, 317, 442
\bibitem[Bouch\'{e}(2001)]{bouche01} Bouch\'{e}, N.,
    Lowenthal, J. D., Bershady, M. A., Churchill, C. W., \&
    Steidel, C. C. 2001, ApJ, 550, 585
\bibitem[Browning et al.(2003)]{browning03} Browning, M. K.,
    Tumlinson, J., \& Shull, J. M. 2003, ApJ, 582, 810
\bibitem[Bunker et al.(1999)]{bunker99} Bunker, A. J.,
    Warren, S. J., Clements, D. L., Williger, G. M., \&
    Hewett, P. C. 1999, MNRAS, 309, 875
\bibitem[Burbidge et al.(1996)]{burbidge96} Burbidge, E. M.,
    Beaver, E. A., Cohen, R. D., Junkkarinen, V. T., \&
    Lyons, R. W. 1996, AJ, 112, 2533
\bibitem[Cazaux \& Tielens(2002)]{cazaux02} Cazaux, S., \&
    Tielens, A. G. G. M. 2002, ApJ, 575, L29
\bibitem[Cen et al.(2003)]{cen03} Cen, R., Ostriker, J. P., Prochaska,
    J. X., \& Wolfe, A. M. 2003, ApJ, 598, 741
\bibitem[Chengalur \& Kanekar(2000)]{chengalur00} Chengalur, J. N., \&
    Kanekar, N. 2000 MNRAS, 318, 303
\bibitem[Cooke et al.(1997)]{cooke97} Cooke, A. J., Espey, B.,
    \& Carswell, R. F. 1997, MNRAS, 284, 552
\bibitem[Curran et al.(2004)]{curran04} Curran, S. J., Webb, J. K.,
    Murphy, M. T., \& Carswell, R. F. 2004, MNRAS, 351, L24
\bibitem[Draine \& Bertoldi(1996)]{draine96} Draine, B. T., \&
    Bertoldi, F. 1996, ApJ, 468, 269
\bibitem[Draine \& Lee(1984)]{draine84} Draine, B. T., \&  Lee, H. M.
    1984, ApJ, 285, 89
\bibitem[Draptz et al.(1977)]{drapatz77} Drapatz, S., \&
    Michel, K. W. 1977, A\&A, 56, 353
\bibitem[Ellison et al.(2002)]{ellison02} Ellison, S. L.,
    Yan, L., Hook, I. M., Pettini, M., Wall, J. V., \& Shaver, P.
    2002, A\&A, 383, 91
\bibitem[Fall et al.(1989)]{fall89} Fall, S. M, Pei, Y. C. \& 
    McMahon, R. G. 1989, ApJ, 341, L5
\bibitem[Ferrara et al.(1999)]{ferrara99} Ferrara, A., Nath, B.,
    Sethi, S. K., \& Shchekinov, Y. 1999, MNRAS, 303, 301
\bibitem[Gardner et al.(2001)]{gardner01} Gardner, J. P.,
    Katz, N., Hernquist, L., \& Weinberg, D. H. 2001,
    ApJ, 559, 131
\bibitem[Ge \& Bechtold(1997)]{ge97} Ge, J., \& Bechtold, J. 1997,
    ApJ, 477, L73
\bibitem[Ge et al.(2001)]{ge01} Ge, J., Bechtold, J., \& Kulkarni,
    P. 2001, ApJ, 547, L1
\bibitem[Giallongo et al.(1996)]{giallongo96} Giallongo, E.,
    Cristiani, S., D'Odorico, S., Fontana, A., \& Savaglio, S. 1996,
    ApJ, 466, 46
\bibitem[Gry et al.(2002)]{gry02} Gry, C., Boulanger, F.,
    Nehm\'{e}, C., Pineau de For\^{e}ts, G., Habart, E., \&
    Falgarone, E. 2002, A\&A, 391, 675
\bibitem[Haardt \& Madau(1996)]{haardt96} Haardt, F. \& Madau, P.
    1996, ApJ, 461, 20
\bibitem[Habing(1968)]{habing68} Habing, H. J. 1968, Bull.\ Astr.\
    Inst.\ Netherlands, 19, 421
\bibitem[Haehnelt et al.(1998)]{haehnelt98} Haehnelt, M. G., Steinmetz,
    M., \& Rauch, M. 1998, ApJ, 495, 647
\bibitem[Heithausen(2004)]{heithausen04} Heithausen, A. 2004,
    ApJ, 606, L13
\bibitem[Hirashita et al.(2003)]{hirashita_buat03} Hirashita, H.,
    Buat, V., \& Inoue, A. K. 2003, A\&A, 410, 83
\bibitem[Hirashita \& Ferrara(2002)]{hirashita02} Hirashita, H., \&
    Ferrara, A. 2002, MNRAS, 337, 921
\bibitem[Hirashita et al.(2003)]{hirashita03} Hirashita, H.,
    Ferrara, A., Wada, K., \& Richter, P. 2003, MNRAS, 341, L18
\bibitem[Hirashita \& Hunt(2004)]{hirashita04} Hirashita, H., \&
    Hunt, L. K. 2004, A\&A, 421, 555
\bibitem[Hollenbach \& McKee(1979)]{hollenbach79} Hollenbach, D. J.,
    \& McKee, C. F. 1979, ApJS, 41, 555
\bibitem[Iglesias-P\'{a}ramo et al.(2004)]{iglesias04}
    Iglesias-P\'{a}ramo, J., Buat, V., Donas, J., Boselli, A.,
    \& Milliard, B. 2004, A\&A, 419, 109
\bibitem[Jura(1974)]{jura74} Jura, M. 1974, ApJ, 191, 375
\bibitem[Jura(1975)]{jura75} Jura, M. 1975, ApJ, 197, 581
\bibitem[Kulkarni et al.(2000)]{kulkarni00} Kulkarni, V. P.,
    Hill, J. M., Schneider, G., Weymann, R. J.,
    Storrie-Lombardi, L. J., Rieke, M. J., Thompson, R. I., \&
    Jannuzi, B. T. 2000, ApJ, 536, 36
\bibitem[Lanzetta et al.(1989)]{lanzetta89} Lanzetta, K. M.,
    Wolfe, A. M., \& Turnshek, D. A. 1989, ApJ, 344, 277
\bibitem[Lanzetta et al.(1995)]{lanzetta95} Lanzetta, K. M., Wolfe,
    A. M., \& Turnshek, D. A. 1995, ApJ, 440, 435
\bibitem[Ledoux et al.(2002)]{ledoux02} Ledoux, C., Srianand, R., \&
    Petitjean, P. 2002, A\&A, 392, 781
\bibitem[Ledoux et al.(2003)]{ledoux03} Ledoux, C., Petitjean, P.,
    \& Srianand, R. 2003, MNRAS, 346, 209
\bibitem[Levshakov et al.(2002)]{levshakov02} Levshakov, S. A.,
    Dessauges-Zavadsky, M., D'Odorico, S., \& Molaro, P. 2002, ApJ,
    565, 696
\bibitem[Levshakov et al.(2000)]{levshakov00} Levshakov, S. A.,
    Molaro, P., Centuri\'{o}n, M., D'Odorico, S., Bonifacio, P.,
    \& Vladilo, G. 2000, A\&A, 361, 803
\bibitem[Liszt(2002)]{liszt02} Liszt, H. 2002, A\&A, 389, 393
\bibitem[Lu et al.(1996)]{lu96} Lu, L., Sargent, W. L. W.,
    Barlow, T. A., Churchill, C. W., \& Vogt, S. S. 1996, ApJS,
    107, 475
\bibitem[Marggraf et al.(2004)]{marggraf04} Marggraf, O.,
    Bluhm, H., \& de Boer, K. S. 2004, A\&A, 416, 251
\bibitem[McKee \& Ostriker(1977)]{mckee77} McKee, C. F., \&
    Ostriker, J. P. 1977, ApJ, 218, 148
\bibitem[M{\o}ller et al.(2002)]{moller02} M{\o}ller, P.,
    Warren, S. J., Fall, S. M., Fynbo, J. U., \&
    Jakobsen, P. 2002, ApJ, 574, 51
\bibitem[Murphy \& Liske(2004)]{murphy04} Murphy, M. T., \&
    Liske, J. 2004, MNRAS, submitted
\bibitem[Nagamine et al.(2004)]{nagamine04} Nagamine, K.,
    Springel, V., \& Hernquist, L. 2004, MNRAS, 348, 435
\bibitem[Okoshi et al.(2004)]{okoshi04} Okoshi, K.,
    Nagashima, M., Gouda, N., \& Yoshioka, S. 2004, ApJ, 603, 12
\bibitem[Omukai(2000)]{omukai00} Omukai, K. 2000, ApJ, 534, 809
\bibitem[Pei \& Fall(1995)]{pei95} Pei, Y. C., \& Fall, S. M.
    1995, ApJ, 454, 69
\bibitem[Pei, Fall, \& Hauser(1999)]{pei99} Pei, Y. C.,
    Fall, S. M., \& Hauser, M. G. 1999, ApJ, 522, 604
\bibitem[P\'{e}roux et al.(2003)]{peroux03} P\'{e}roux, C.,
    McMahon, R. G., Storrie-Lombardi, L. J., \& Irwin, M. J.
    2003, MNRAS, 346, 1103
\bibitem[Petitjean et al.(2000)]{petitjean00} Petitjean, P.,
    Srianand, R., \& Ledoux, C. 2000, A\&A, 364, L26
\bibitem[Petitjean et al.(2002)]{petitjean02} Petitjean, P.,
    Srianand, R., \& Ledoux, C. 2002, MNRAS, 332, 383
\bibitem[Pettini et al.(1999)]{pettini99} Pettini, M., Ellison,
    S. L., Steidel, C. C., \& Bowen, D. V. 1999, ApJ, 510, 576
\bibitem[Pettini et al.(1994)]{pettini94} Pettini, M. H.,
    Smith, L. J., Hunstead, R. W., \& King, D. L. 1994, ApJ, 426,
    79
\bibitem[Prochaska \& Wolfe(2002)]{prochaska02} Prochaska, J. X., \&
    Wolfe, A. M. 2002, ApJ, 566, 68
\bibitem[Quast et al.(2002)]{quast02} Quast, R., Baade, R., \&
    Reimers, D. 2002, A\&A, 386, 796
\bibitem[Rao et al.(2003)]{rao03} Rao, S. M., Nestor, D. B.,
    Turnshek, D. A., Lane, W. M., Monier, E. M., \& Bergeron, J.
    2003, ApJ, 595, 94
\bibitem[Reimers et al.(2003)]{reimers03} Reimers, D., Baade, R.,
    Quast, R., \& Levshakov, S. A. 2003, A\&A, 410, 785
\bibitem[Richter 2000]{richter2000}
    Richter, P., 2000, A\&A, 359, 1111
\bibitem[Richter et al.(2003a)]{richter03a}
    Richter, P., Sembach, K. R., \& Howk, J. C. 2003, A\&A,
    405, 1013
\bibitem[Richter et al.(2003b)]{richter03b}
    Richter, P., Wakker, B. P., Savage, B. D., Sembach, K. R.,
    2003, ApJ, 586, 230
\bibitem[Salucci \& Persic(1999)]{salucci99} Salucci, P., \&
    Persic, M. 1999, MNRAS, 309, 923
\bibitem[Schaye(2001)]{schaye01} Schaye, J. 2001, ApJ, 562, L95
\bibitem[Schaye(2004)]{schaye04} Schaye, J. 2004, ApJ, submitted
\bibitem[Scott et al.(2000)]{scott00} Scott, J., Bechtold, J.,
    Dobrzycki, A., \& Kulkarni, V. P. 2000, ApJS, 130, 67
\bibitem[Scott et al.(2002)]{scott02} Scott, J., Bechtold, J.,
    Morita, M., Dobrzycki, A., \& Kulkarni, V. P. 2002, ApJ, 571,
    665
\bibitem[Shibai et al.(1999)]{shibai99} Shibai, H., Okumura, K.,
    \& Onaka, T. 1999, in T. Nakamoto, ed., Star Formation 1999,
    Nobeyama, Nobeyama Radio Observatory, p.\ 67
\bibitem[Shibai et al.(2001)]{shibai01} Shibai, H.,
    Takeuchi, T. T., Rengarajan, T. N., \& Hirashita, H. 2001,
    PASJ, 53, 589
\bibitem[Smith et al.(1978)]{smith78} Smith, L. F., Biermann, P.,
    \& Mezger, P. G. 1978, A\&A, 66, 65
\bibitem[Spitzer(1978)]{spitzer78} Spitzer, L., Jr.\ 1978,
    Physical Processes in the Interstallar Medium, Wiley,
    New York
\bibitem[Spitzer \& Zweibel(1974)]{spitzer74} Spitzer, L., Jr.,
    \& Zweibel, E. G. 1974, ApJ, 191, L127
\bibitem[Srianand et al.(2000)]{srianand00} Srianand, R.,
    Petitjean, P., \& Ledoux, C. 2000, Nat, 408, 931
\bibitem[Storrie-Lombardi \& Wolfe(2002)]{storrie02}
    Storrie-Lombardi, L. J., \& Wolfe, A. 2000, ApJ, 543, 552
\bibitem[Takeuchi et al.(2003)]{takeuchi03} Takeuchi, T. T.,
    Hirashita, H.,
    Ishii, T. T., Hunt, L. K., \& Ferrara, A. 2003, MNRAS,
    343, 839
\bibitem[Tumlinson et al.\,2002]{tumlinson2002}
    Tumlinson, J., et al., 2002, ApJ, 566, 857
\bibitem[Varshalovich et al.(2001)]{varshalovich01}
    Varshalovich, D. A., Ivanchik, A. V., Petitjean, P.,
    Srianand, R., \& Ledoux, C. 2001, Astron.\ Lett., 27, 683
\bibitem[Vladilo(2002)]{vladilo02} Vladilo, G. 2002, A\&A, 391, 407
\bibitem[Wada \& Norman(2001)]{wada01} Wada, K., \& Norman, C. A.
    2001, ApJ, 546, 172
\bibitem[Wolfe et al.(2003a)]{wolfe03a} Wolfe, A. M.,
    Prochaska, J. X., \& Gawiser, E. 2003a, ApJ, 593, 215
\bibitem[Wolfe et al.(2003b)]{wolfe03b} Wolfe, A. M.,
    Gawiser, E., \& Prochaska, J. X. 2003b, ApJ, 593, 235
\bibitem[Wolfe et al.(1986)]{wolfe86} Wolfe, A. M., Turnshek, D. A.,
    Smith, H. E., \& Cohen, R. D. 1986, ApJ, 61, 249
\bibitem[Wolfire et al.(1995)]{wolfire95} Wolfire, M. G.,
    Hollenbach, D., McKee, C. F., Tielens, A. G. G. M., \&
    Bakes, E. L. O. 1995, ApJ, 443, 152
\bibitem[Zuo et al.(1997)]{zuo97} Zuo, L., Beaver, E. A.,
    Burbidge, E. M., Cohen, R. D., Junkkarinen, V. T., \&
    Lyons, R. W. 1997, ApJ, 477, 568
\end{thebibliography}
\end{document}